\documentclass[11pt,sort&compress]{elsarticle}
\usepackage{amsmath,amssymb,epsfig,multirow,xcolor,hyperref}

\begin{document}

\title{
Optimized Parallelization of Boundary Integral Poisson-Boltzmann Solvers 
}
\author[smu]{Xin Yang}
\ead{xiny@smu.edu}

\author[smu]{Elyssa Sliheet}
\ead{esliheet@smu.edu}

\author[smu]{Reece Iriye}
\ead{ririye@smu.edu}

\author[smu]{Daniel Reynolds}
\ead{reynolds@smu.edu}

\author[smu]{Weihua~Geng\corref{cor1}}
\ead{wgeng@smu.edu}

\cortext[cor1]{Corresponding author}

\address[smu]{Department of Mathematics, Southern Methodist University, Dallas, TX 75275 USA}

\begin{abstract}
\noindent
The Poisson-Boltzmann (PB) model governs the electrostatics of solvated biomolecules, 
i.e., potential, field, energy, and force. 
These quantities can provide useful information about protein properties, functions, and dynamics. 
By considering the advantages of current algorithms and computer hardware, 
we focus on the parallelization of the treecode-accelerated boundary integral (TABI) PB solver 
using the Message Passing Interface (MPI) on CPUs 
and the direct-sum boundary integral (DSBI) PB solver 
using KOKKOS on GPUs. 
We provide optimization guidance for users 
when the DSBI solver on GPU or the TABI solver with MPI on CPU should be used 
depending on the size of the problem.  
Specifically, when the number of unknowns is smaller than a predetermined threshold, 
the GPU-accelerated DSBI solver converges rapidly thus has the potential to perform PB model-based molecular dynamics or Monte Carlo simulation. 
As practical appliations, our parallelized boundary integral PB solvers are used to solve electrostatics
on selected proteins that play significant roles 
in the spread, treatment, and prevention of COVID-19 virus diseases. 
For each selected protein, the simulation produces the electrostatic solvation energy 
as a global measurement and electrostatic surface potential for local details.

\end{abstract}


\maketitle


\section{Introduction}
\label{intro}
\noindent
Since the invention of X-ray crystallography and nuclear magnetic resonance (NMR) spectroscopy, the structures of biomolecules such as proteins and nucleic acids has become largely available to computational bioscientists. For example, as of October 13, 2023, there are 210,554 structures stored in the protein data bank (PDB) \cite{PDB}. 
These available structures make the computational simulation of biomolecules possible, which tremendously promotes the advancement of structural biology research. 
At the molecular level, where length is normally measured in angstroms (\AA), the structures, properties, functions, and dynamics of proteins are critically determined by the electrostatic interactions between proteins, ligands, and their solvent environment. 
The fundamental physical theory characterizing these electrostatics interactions is Gauss's law, 
the differential form of which leads to the 
Poisson-Boltzmann (PB) model that incorporates quantities such as 
media permittivity, 
protein charge distribution, 
electrolytes distribution in solvent, temperature, etc. 
With the assistance of numerical algorithms and computational tools, 
the PB model has broad applications in biomolecular simulations, including 
protein structure \cite{Cherezov:2007}, 
protein-protein interaction \cite{Dong:2003,Huang:2012}, 
chromatin packing \cite{Beard:2001}, 
protein pKa values, 
protein-membrane interactions,  
\cite{Zhou:2010,Callenberg:2010}, 
binding energy \cite{Nguyen:2017,Wilson:2022}, 
solvation free energy\cite{Simonson:2002,Wagoner:2006},
ion channel profiling \cite{Unwin:2005}, etc.

Solving the PB model accurately and efficiently 
is challenging due to a variety of factors, such as 
the large problem dimension, 
the complex geometry of the protein, 
the jump of dielectric constants across media interface, and
the singular charge representation.
In the past 15 years, we have been working on developing fast and accurate PB solvers as useful tools for theoretical and computational bio-scientists \cite{Geng:2007,Geng:2013,Geng:2013a,Geng:2013b,Geng:2017,Chen:2023}. Among the many attempted approaches, our favorite choices are the boundary integral PB solvers 
\cite{Geng:2013b,Geng:2013a}, 
which potentially have the best combination of efficiency, 
accuracy, 
memory usage, 
and parallelization. 
This manuscript will focus on the parallelization development of boundary integral PB solvers.

Boundary integral PB solvers are amenable to parallel computing because 
solving the boundary integral PB model is essentially similar to 
computing the pairwise interactions 
between the charges or induced charges located at the boundary elements, 
which is in fact an $n$-body problem. The parallelization of these pairwise interactions with $O(n^2)$ computational cost is straight-forward.  
When fast algorithms, such as treecode \cite{Li:2009} or fast mulitpole method (FMM) \cite{Greengard:1987} are involved, extra work is needed to ensure the parallel efficiency \cite{Chen:2021a,Zhang:2015}. This article emphasizes the comparison for parallelization between treecode and direct-sum when different computing hardware are available.   
With MPI implementation using multicore CPUs, for moderate numbers of particles we can build the tree on every CPU/task thus all particle-tree interactions can be concurrently done with high parallel efficiency \cite{Geng:2013b,Chen:2021a}. When the number of particles are huge thus replicating the tree on each task is not possible, domain decomposition approach can be used instead \cite{Salmon:1986,Chen:2021a}.

In recent years, Graphic Processing Units (GPUs) have revolutionized computation-intensive tasks, including high performance computing, call center, autopilot, artificial intelligence, etc. 
One GPU card can consist of hundreds to thousands of cores and can rapidly implement large quantities of tasks, particularly \emph{single instruction multiple data} (SIMD) tasks that benefit from large-scale concurrency. 
The $n$-body problem in which 
all particles concurrently interact with all other particles is thus very suitable for GPU computation. Calculation of $n$-body interactions using direct-sum on GPUs can be traced back to the first decade of the 21st century \cite{Elsen:2006,Nyland:2009}. 
Later, when fast algorithms, such as treecode, were designed for the $n$-body problem, 
their GPU implementation is also possible \cite{Hamada:2009,Bedorf:2012,Burtscher:2011}. 
However, these complex algorithms are challenging to implement on GPUs, thereby reducing the achievable level of parallel 
speedup. For example, as shown in \cite{Belleman:2008}, when more than 100k particles are involved in an $n$-body interaction, an NVIDIA GeForce 8800GTX GPU outperforms an Intel Xeon CPU running at 3.4 GHz by a factor of about 100 using direct summation as opposed to a factor of about 10 using Barnes-Hut treecode \cite{Barnes:1986}. 

Our argument is that when a fast algorithm is available to replace the direct method e.g.~treecode vs.~direct summation, 
there is a break-even number $n_b$ such that the fast algorithm should only be applied when the problem size exceeds this threshold. 
For example, for an $n$-body problem, consider the $c_1n\log n$ treecode algorithm against the $c_2n^2$ direct sum, where $c_1$ and $c_2$ are some constants calculated from the algorithms, the break-even number $n_b$ is the $n$ that satisfies the equation $c_1 n\log n = c_2n^2$. 
Due to the algorithmic simplicity of direct summation, the break-even number $n_b$ becomes considerately larger on GPUs in comparison to CPUs. 
Thus when the efficiency of the implementation is of critical importance, 
such as when these are used repeatedly within molecular dynamics or Monte Carlo simulations, 
we should use direct summation on GPU when the problem size is smaller than $n_b$.
As we will show in the section of numerical results, $n_b$ for the current GPU/CPU hardware conditions used for this project is about 250,000. In practice, molecular surface with 250,000 triangular elements can 
readily represent a large group of proteins with less than 2000-3000 atoms. 
For larger problems, we can use     
the MPI-based parallel treecode \cite{Chen:2021a} 
or use the domain decomposition approach \cite{Salmon:1986}. 
In short, every method has its own advantageous size of problems. 
We also note that the previous studies that have utilized a mix of hardware and methods, including 
the elegant work using MPI and GPUs \cite{Vaughn:2020a}.  We additionally point out the study by Wilson et al.\cite{Wilson:2021} wherein  
an Orthogonal Recursive Bisection (ORB) tree based domain decomposition \cite{Salmon:1986} is applied to allow each MPI task to only contain a local essential tree. 
The majority of the computation is then transferred to the GPU available on the node. This work is efficient for large problems but requires additional hardware support such that each node should have compatible multiple CPUs and a GPU card. 





In this work we focus on two approaches for the parallelization of boundary integral PB solvers. 
One is the parallelization of the treecode-accelerated boundary integral (TABI) solver using MPI which builds an identical tree on each task/CPU. 
Its parallelization occurs at four stages of the TABI solver: 
source term computation, 
treecode for matrix-vector product, 
preconditioning, and
energy computation. 
Among these stages, 
we apply the schemes developed for $n$-body parallelization \cite{Chen:2021a} to a more complicated boundary integral PB problem, 
and develop MPI-based parallelization for the preconditioning scheme designed for boundary integral solvers \cite{Chen:2018}. 
Our second parallelization approach focuses on GPU-based parallelization of the 
direct-sum boundary integral (DSBI) solver, 
which concurrently computes the source term, matrix-vector product, and energy computation. 
We note that we did not use the preconditioning scheme in this approach 
since its GPU implementation is complicated and inefficient. 
As we will explain in the section of theories and algorithms and show in the section of numerical results, 
for some proteins whose surface triangulations contain a few poor-quality triangles, without using the preconditioning schemes will take longer time for the GMRES \cite{Saad:1986} iterative solver to converge, which is a limitation we are still working to improve. 
%

The rest of the paper is organized as follows. 
In Section \ref{method}, we introduce the theories and algorithms involving 
the PB model, 
the boundary integral method, 
the treecode algorithm, 
the GMRES preconditioning scheme, 
the GPU and MPI parallelization,
and biological information on the COVID-19 related proteins. 
Following that, we provide numerical results and related discussion in Section \ref{numericalresults}. 
This paper ends with short sections on software dissemination and concluding remarks.

\section{Theories and algorithms}\label{method}

In this section, we will first briefly describe the PB model. 
Then we present a brief introduction to the treecode-accelerated boundary integral (TABI) PB solver. 
After that, we move to the parallelization schemes, 
which include a MPI based TABI solver 
and 
a GPU-accelerated direct-sum boundary integral (DSBI) Solver. 
This section ends with the introduction of the proteins that 
are involved in COVID-19 diseases, and are the target for our simulations.

\subsection{The PB model}
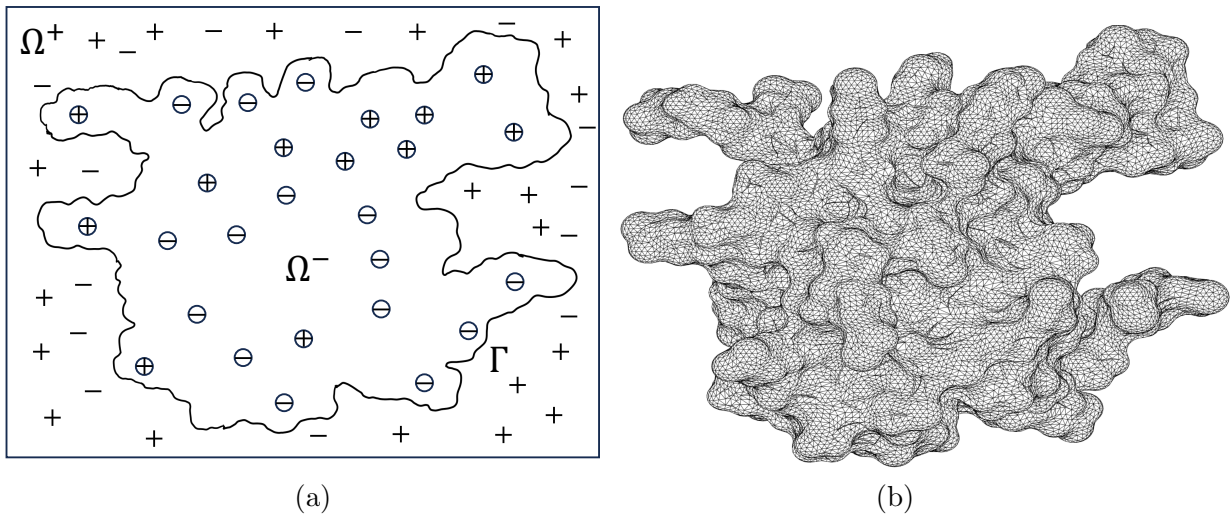
\begin{figure}[!tbp]
\setlength{\unitlength}{1cm}
\begin{picture}(5,1)
\put( 2.8, -3){\Large $\Omega^-$}
\put( 0.5, -1.7){\Large $\Omega^+$}
\put( 4, -2.8){\Large $\Gamma$}
\end{picture}
\begin{center}
\includegraphics[width=3.2in]{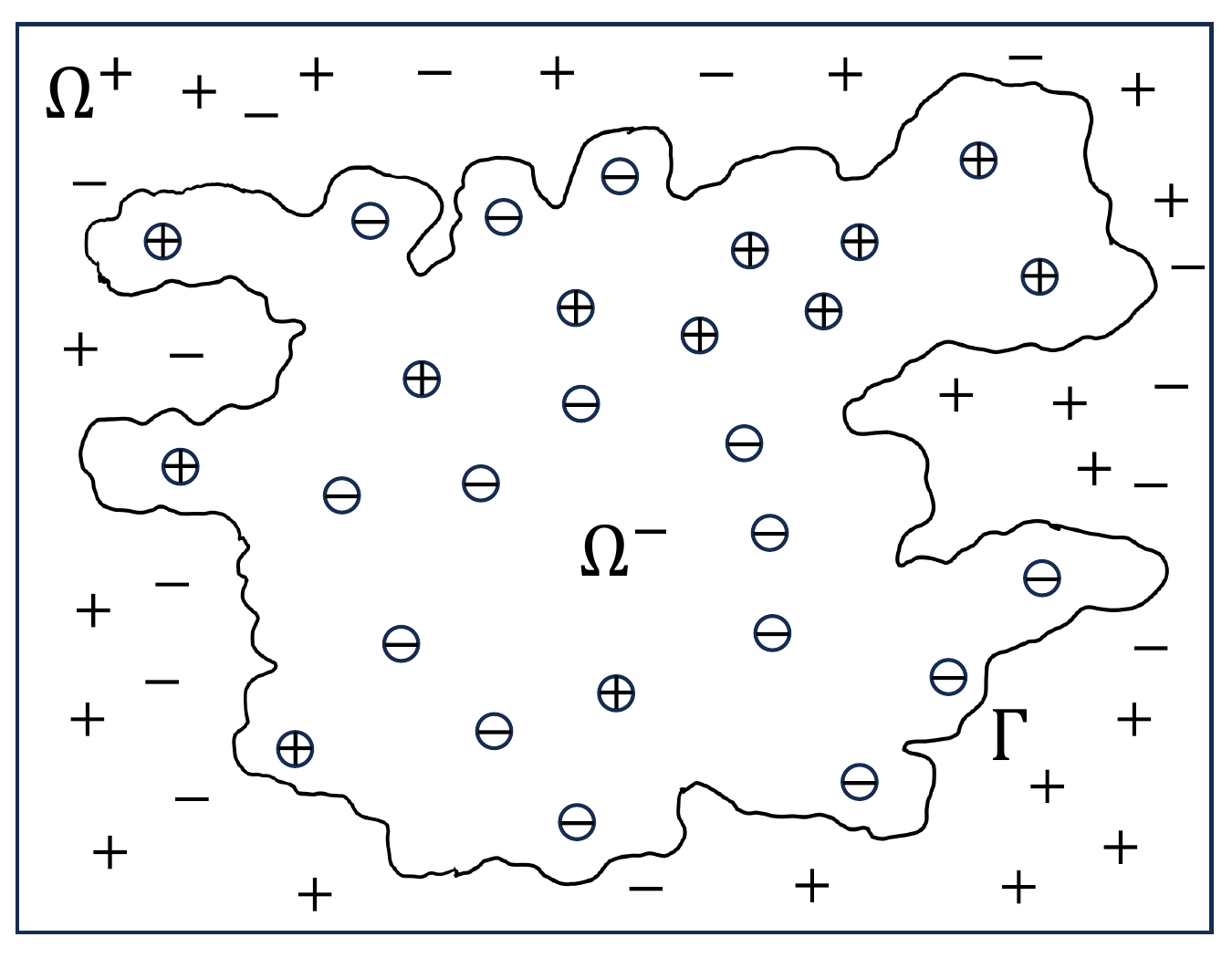}
\includegraphics[width=3.2in]{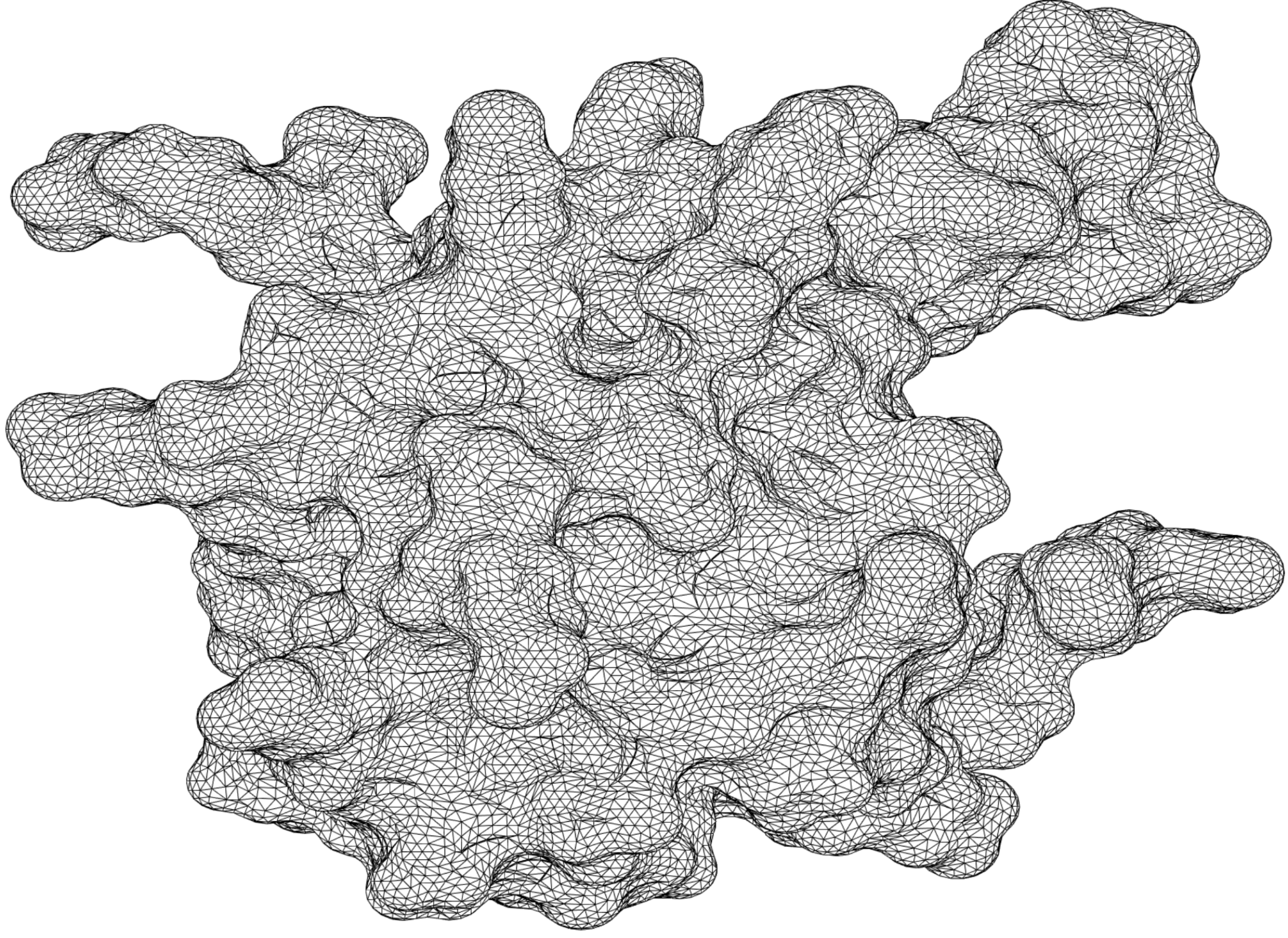}\\
\hskip -0.1in (a) \hskip 2.8in (b) 
\caption{(a) The PB model with molecular domain $\Omega^-$ and solvent domain $\Omega^+$ separated by the molecular surface $\Gamma$, (b) The triangulation surface of protein 6yi3, whose 2-d illustration is used in (a) as the molecular surface $\Gamma$.}
\label{fig_1}
\end{center}
\end{figure}

The PB model is illustrated in Fig.~\ref{fig_1}(a). 
Consider a domain $\Omega$ in $\mathbb{R}^3$ 
divided by the molecule surface $\Gamma$ 
into the molecule domain $\Omega^-$ with dielectric constant $\epsilon^-$ 
and the solvent domain $\Omega^+$ with dielectric constant $\epsilon^+$, 
i.e., $\Omega=\Omega^-\bigcup\Omega^+$. 
Note that the surface $\Gamma$ here is a 2-d illustration of a 3-d molecular surface of the protein 6yi3 whose triangulated molecular surface is given in 
Fig.~\ref{fig_1}(b). 
This triangulated surface provides the discretized space for the boundary integral method that we adopt. 
Note that all the proteins used in this article are from the Protein Data Bank (PDB)\cite{PDB}. We refer to the proteins using their 4-digit PDB ID numbers, and detailed information on each can be found on the PDB website.  
Protein 6yi3 is one of the COVID-19 proteins of interest in this study, which will be described in the following sections. 

Charges in $\Omega^-$ illustrated as circled ``+" and ``-" signs are partial charges assigned to the centers of atoms by using the force field, 
while charges in $\Omega^+$ illustrated as ``+" and ``-" signs are mobile ions described by the Boltzmann distribution. 
Assuming equal amounts of positively and negatively charged electrolytes, then according to Gauss's law the charge distribution for ${\bf r} \in \Omega^+$ 
\begin{equation}
-\nabla\cdot(\epsilon({\bf r}) \nabla \phi({\bf r}))+\bar{\kappa}^2({\bf r})\sinh{(\phi({\bf r}))=\rho({\bf r})},
\label{eqNPBE}
\end{equation}
subject to interface continuity for the potential $\phi$ and flux density $\epsilon \phi_{\bf n}$,
\begin{equation}
[\phi]_\Gamma=0 \text{ and } [\epsilon \phi_{\nu}]_\Gamma=0.
\label{eqJump}
\end{equation}
Here ${\nu}=(n_x,n_y,n_z)$ is the outward normal direction of the
interface $\Gamma$, 
$\phi_{\nu}=\frac{\partial \phi }{\partial {\nu}}$ is the directional derivative in ${\nu}$,
and the notation $[f]_\Gamma=f^+-f^-$ is the difference of the function $f$ cross the interface $\Gamma$. 

In Eqs.~(\ref{eqNPBE}) and (\ref{eqJump}), $\epsilon$ is a piecewise function for the dielectric constants in $\Omega^-$ and $\Omega^+$.  
Here $\kappa$ is the inverse Debye 
screening length, which measures ionic strength and its modified version $\bar{\kappa}$ is given as $\bar{\kappa}^2=\epsilon^+\kappa^2$. The value of $\kappa$ is nonzero in $\Omega^+$ only. 
In our numerical implementation, the unit of length is {\AA} and the potential $\phi$ has units $e_c/$\AA~.
To compute the free energy in kcal/mol/$e_c$, after computing $\phi$ it must be multiplied by a factor of $332.0716$. 
For further details regarding the definitions of variables, coefficients, and units in the PB model we refer the reader to \cite{Geng:2015,Holst:1994}.  

The source term $\rho$ in Eq.~(\ref{eqNPBE}) is the summation of the charge distribution in $\Omega^-$ using delta functions for $N_c$ partial charges located at ${\bf r}_i$, i.e.,
$\rho({\bf r})=4\pi C\sum\limits_{i=1}^{N_c} q_{i}\delta({\bf r}-{\bf r}_i)$, where $C$ is a constant to balance the units. 
Due to the singular partial charges in the source $\rho({\bf r})$, the electrostatic potential $\phi({\bf r})$ goes to infinity as ${\bf r} \to {\bf r}_i$.

From the description of the PB model above, it is clear that numerical solutions to the PB model face many challenges, including the complex geometry as illustrated in Fig.~\ref{fig_1}(b), interface conditions as in Eq.~\ref{eqJump}, the singular charges in $\rho$, the infinite computational domain in $\mathbb{R}^3$, etc. These numerical difficulties are challenging for grid-based methods, but can be conveniently addressed with the boundary integral method as described next.

\subsection{Boundary Integral PB solvers}

In this subsection we describe the boundary integral method
for computing the electrostatic surface potential and solvation free energy~\cite{Geng:2013b,Geng:2013}.
We present 
the boundary integral form of the PB implicit solvent model, 
the discretization of the boundary integral equations,
the treecode algorithm for accelerating the matrix-vector product, 
and the preconditioning scheme to alleviate the rising condition number
when the triangulation quality deteriorates due to complex geometry. 
Note these algorithms were described in previous work \cite{Li:2009,Geng:2013b,Chen:2018} and this paper emphasize the parallelization of them using multicore CPUs and GPUs. 
Following the tradition of the boundary integral method, we use ${\bf x}$ and ${\bf y}$ to represent the spatial position as opposed to the previously used ${\bf r}$. We also denote $\Omega_1=\Omega^-$, $\Omega_2=\Omega^+$, $\epsilon_1=\epsilon^-$, and $\epsilon_2=\epsilon^+$.

\subsubsection{Boundary integral form of PB model}

This section summarizes the well-conditioned boundary integral form of the 
PB implicit solvent model we employ~\cite{Juffer:1991,Geng:2013b}.
Applying Green's second identity and properties of fundamental solutions 
to Eq.~(\ref{eqNPBE}) yields the electrostatic potential in each domain,
\begin{subequations}
\begin{align}
\label{eqbim_1}
\phi({\bf x}) = 
&\int_\Gamma \left[G_0({\bf x},
{\bf y})\frac{\partial\phi({\bf y})}{\partial{\nu}} -\frac{\partial G_0({\bf x}, {\bf y})}{\partial{\nu}_{{\bf y}}} \phi({\bf y})
\right]\mathrm dS_{{\bf y}} + \sum_{k=1}^{N_{c}}q_k G_0({\bf x}, {\bf y}_k), \quad {\bf x} \in \Omega_1, \\
\label{eqbim_2}
\phi({\bf x}) = 
&\int_\Gamma\left[-G_\kappa({\bf x},
{\bf y})\frac{\partial\phi({\bf y})}{\partial{\nu}} + \frac{\partial G_\kappa({\bf x}, {\bf y})}{\partial{\nu}_{{\bf y}}}
\phi({\bf y})\right]\mathrm dS_{{\bf y}}, \quad {\bf x} \in \Omega_2,
\end{align}
\end{subequations}
where $G_0({\bf x},{\bf y})$ and $G_\kappa({\bf x}, {\bf y})$ are the 
Coulomb and screened Coulomb potentials, respectively
\begin{equation}
G_0({\bf x}, {\bf y}) = \frac{1}{4\pi|{\bf x}-{\bf y}|}
\quad\text{and}\quad
\label{eq_potential}
G_\kappa({\bf x}, {\bf y}) = \frac{e^{-\kappa|{\bf x}-{\bf y}|}}{4\pi|{\bf x}-{\bf y}|}.
\end{equation}
Then applying the interface conditions in Eq.~(\ref{eqJump}) 
with the differentiation of electrostatic potential in each domain
yields 
a set of boundary integral equations relating the
surface potential $\phi_1$
and
its normal derivative $\partial\phi_1/\partial{\nu}$ on $\Gamma$,
\begin{subequations}
\begin{align}
\label{eqbim_3}
\frac{1}{2}\left(1+\varepsilon\right)\phi_1({\bf x}) & =
\int_\Gamma \left[K_1({\bf x}, {\bf y})\frac{\partial\phi_1({\bf y})}{\partial{\nu}} +
K_2({\bf x}, {\bf y})\phi_1({\bf y})\right]\mathrm dS_{{\bf y}}+S_{1}({\bf x}),
\quad {\bf x}\in\Gamma, \\
\label{eqbim_4}
\frac{1}{2}\left(1+\frac{1}{\varepsilon}\right)\frac{\partial\phi_1({\bf x})}{\partial{\nu}} & =
\int_\Gamma \left[K_3({\bf x}, {\bf y})\frac{\partial\phi_1({\bf y})}{\partial{\nu}} +
K_4({\bf x}, {\bf y})\phi_1({\bf y})\right]\mathrm dS_{{\bf y}}
+S_{2}({\bf x}),
\quad {\bf x} \in \Gamma,
\end{align}
\end{subequations}
where $\varepsilon = \varepsilon_2/\varepsilon_1$.
As given in Eqs.~(\ref{Eq_K1}-\ref{Eq_K2}) and (\ref{source_terms}), the kernels $K_{1,2,3,4}$ and source terms $S_{1,2}$ are linear combinations of $G_0$, $G_k$, 
and their first and second order normal derivatives~\cite{Juffer:1991,Geng:2013b},
\begin{subequations}
\begin{align} 
\label{Eq_K1}
K_1({\bf x}, {\bf y}) = 
&\,{G_{0}({\bf x},{\bf y})}-{G_{\kappa}({\bf x},{\bf y})},
\quad
K_2({\bf x}, {\bf y}) = \varepsilon\frac{\partial G_{\kappa}({\bf x},{\bf y})}{\partial\nu_{{\bf y}}}-\frac{\partial
G_{0}({\bf x},{\bf y})}{\partial\nu_{{\bf y}}}, \\
\label{Eq_K2}
K_3({\bf x}, {\bf y}) = 
&\,\frac{\partial G_{0}({\bf x},{\bf y})}{\partial\nu_{{\bf x}}}-\frac{1}{\varepsilon}\frac{\partial G_{\kappa}({\bf x},{\bf y})}{\partial\nu_{{\bf x}}},
\quad
K_4({\bf x}, {\bf y}) =
\frac{\partial^2 G_\kappa({\bf x},{\bf y})}{\partial\nu_{{\bf x}}\partial\nu_{{\bf y}}}-\frac{\partial^2G_{0}({\bf x},{\bf y})}{\partial\nu_{{\bf x}}\partial\nu_{\bf y}},
\end{align}
\end{subequations}
where the normal derivative with respect to ${\bf x}$ is given by
\begin{align}
\frac{\partial G({\bf x},{\bf y})}{\partial\nu_{{\bf x}}} & =
-\nu({\bf x}) \cdot \nabla_{{\bf x}}G({\bf x},{\bf y}) =
-\sum_{m=1}^3\nu_m({\bf x})\partial_{x_m}G({\bf x},{\bf y}),
\label{Eq_Ds_x}
\end{align}
the normal derivative with respect to ${\bf y}$ is given by
\begin{align}
\frac{\partial G({\bf x},{\bf y})}{\partial\nu_{{\bf y}}} & =
\nu({\bf y}) \cdot \nabla_{{\bf y}}G({\bf x},{\bf y}) =
\sum_{n=1}^3\nu_n({\bf y})\partial_{y_n}G({\bf x},{\bf y}),
\label{Eq_Ds_x}
\end{align}
the second normal derivative with respect to ${\bf x}$ and ${\bf y}$ is given by
\begin{equation}
\label{Eq_Ds_xy}
\frac{\partial G({\bf x},{\bf y})}{\partial\nu_{{\bf y}}\partial\nu_{{\bf x}}} =
-\!\sum_{m=1}^3\sum_{n=1}^3
\nu_m({\bf x})\nu_n({\bf y})\partial_{x_m}\partial_{y_n}G({\bf x},{\bf y}),
\end{equation}
and
the source terms $S_{1,2}$ are
\begin{equation}
S_{1}({\bf x}) = \frac{1}{\varepsilon_1}\sum_{k=1}^{N_{c}}q_kG_{0}({\bf x}, {\bf y}_k)
\quad\text{and}\quad
S_{2}({\bf x}) = \frac{1}{\varepsilon_1}\sum_{k=1}^{N_{c}}q_k
\frac{\partial G_{0}({\bf x},{\bf y}_k)}{\partial\nu_{{\bf x}}}.
\label{source_terms}
\end{equation}

Once the potential and its normal derivative are solved from Eqs.~(\ref{eqbim_3})-(\ref{eqbim_4}), the
potential at any point inside the molecule can be computed via Eq.~(\ref{eqbim_1}), or a numerically more accurate formulation may be used from \cite{Juffer:1991}:
\begin{equation}
\phi_1({\bf x}) =
\int_\Gamma \left[K_1({\bf x}, {\bf y})\frac{\partial\phi_1({\bf y})}{\partial{\nu}} +
K_2({\bf x}, {\bf y})\phi_1({\bf y})\right]\mathrm dS_{{\bf y}}+S_{1}({\bf x}),
\quad {\bf x}\in\Omega_1.
\label{eq_inside}
\end{equation}
With the potential and its normal derivative on $\Gamma$, the electrostatic free energy can be obtained by
\begin{equation}
E^{\rm free} = \frac{1}{2}\sum_{k=1}^{N_c}q_k\phi_1({\bf y}_k) =
\frac{1}{2}\displaystyle \sum\limits_{k=1}^{N_c} q_k\left(
\int_\Gamma \left[K_1({\bf y}_k, {\bf y})\frac{\partial\phi_1({\bf y})}
{\partial\nu}+K_2({\bf y}_k, {\bf y})\phi_1({\bf y})\right]\mathrm dS_{{\bf y}}+S_1({\bf y}_k)\right).
\label{free_energy}
\end{equation}   
The electrostatic solvation free energy can also be obtained by
\begin{equation}
E^{\rm sol} = \frac{1}{2}\sum_{k=1}^{N_c}q_k\phi^{\rm reac}({\bf y}_k) =
\frac{1}{2}\displaystyle \sum\limits_{k=1}^{N_c} q_k
\int_\Gamma \left[K_1({\bf y}_k, {\bf y})\frac{\partial\phi_1({\bf y})}
{\partial\nu}+K_2({\bf y}_k, {\bf y})\phi_1({\bf y})\right]\mathrm dS_{{\bf y}},
\label{solvation_energy}
\end{equation}
where $\phi^{\rm reac}({\bf x}) = \phi({\bf x})-S_1({\bf x})$ is the reaction field potential \cite{Juffer:1991,Geng:2013b}. In our numerical results in Section \ref{numericalresults}, 
we focus on solution of the PB equation and calculation of the electrostatic solvation free energy.

\subsubsection{Discretization of boundary integral equations}

The integrals in Eqs.~(\ref{eqbim_3})-(\ref{eqbim_4}) can be discretized by centroid collocation, which is popular due to its simplicity \cite{Geng:2013b}. Alternatively, it can be discretized using more complicated approaches such as node collocation \cite{Lu:2007a,Wilson:2022a}, curved triangles \cite{Geng:2013}, or Galerkin's method \cite{Chen:2023}, with each resulting in a trade-off between accuracy and efficiency. 
Here we employ the centroid collocation approach. 
 
Letting ${\bf x}_i, i = 1,\ldots,N$ denote the triangle centroids of the $N$ triangular elements, 
the discretized Eqs.~(\ref{eqbim_3})-(\ref{eqbim_4})
have the following form for $i=1,\ldots,N$,
\begin{subequations}
\begin{align}
\label{eqbim_5}
\frac{1}{2}\left(1+\varepsilon\right)\phi_1({\bf x}_i) & =
\sum_{{j=1}\atop{j \ne i}}^{N}\left[
K_1({\bf x}_{i}, {\bf x}_j)\frac{\partial\phi_1({\bf x}_j)}{\partial\nu} +
K_2({\bf x}_{i}, {\bf x}_j)\phi_1({\bf x}_j)\right]\!\Delta s_j + S_{1}({\bf x}_i), \\
\label{eqbim_6}
\frac{1}{2}\left(1+\frac{1}{\varepsilon}\right)\frac{\partial\phi_1({\bf x}_i)}{\partial\nu }& =
\sum_{{j=1}\atop{j \ne i}}^{N}\left[
K_3({\bf x}_{i}, {\bf x}_j)\frac{\partial\phi_1({\bf x}_j)}{\partial\nu} +
K_4({\bf x}_{i}, {\bf x}_j)\phi_1({\bf x}_j)\right]\!\Delta s_j  + S_{2}({\bf x}_i),
\end{align}
\end{subequations}
where $\Delta s_j$ is the area of the $j$th boundary element, and  
the term $j = i$ is omitted from the summation to avoid the kernel singularity.
%
Equations~(\ref{eqbim_5})-(\ref{eqbim_6}) 
represent a linear system $Ax = b$,
where 
${ x }$ contains the surface potentials $\phi_1({\bf x}_i)$
and 
normal derivatives $\frac{\partial\phi_1({\bf x}_i)}{\partial{\nu}}$, 
weighted by the element area $\Delta s_i$,
and
${\bf b}$ contains the source terms $S_1({\bf x}_i)$ and $S_2({\bf x}_i)$.
We solve this system using the generalized minimal residual (GMRES) iterative method,
which requires a matrix-vector product in each step~\cite{Saad:1986}.
Since the matrix is dense,
computing the product by direct summation requires $O(n^2)$ operations,
which is prohibitively expensive when $n$ is large. 
These difficulties can be overcome by fast algorithms for $n$-body computations, such as treecode \cite{Geng:2013b,Wilson:2021} and Fast Multipole Methods \cite{Lu:2007,Chen:2023}. In the next section, we describe how the treecode algorithm is used to accelerate the
matrix-vector product calculation.

\subsubsection{Treecode}

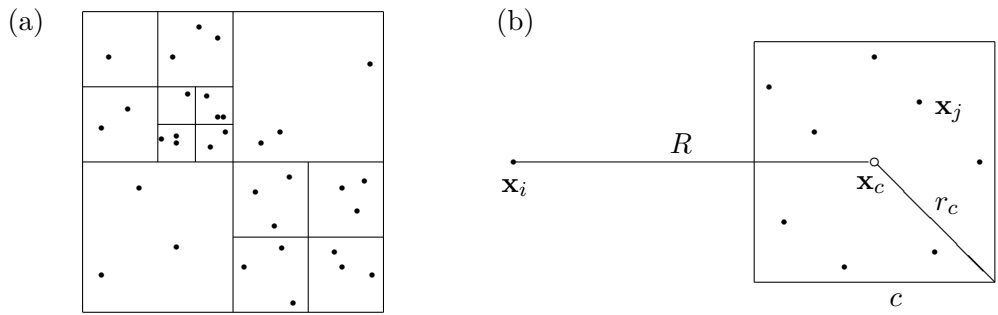
\begin{figure}[htb]
\setlength{\unitlength}{0.5cm}
\hskip 2.5cm
\begin{picture}(28,8)(4,0)

\put( 2.0, 7.5){(a)}
\put( 15.0, 7.5){(b)}

\put(4,0){\line(0,1){8}}
\put(4,0){\line(1,0){8}}
\put(4,8){\line(1,0){8}}
\put(12,0){\line(0,1){8}}

\put(8,0){\line(0,1){8}}
\put(4,4){\line(1,0){8}}

\put(6,4){\line(0,1){4}}
\put(4,6){\line(1,0){4}}

\put(7,4){\line(0,1){2}}
\put(6,5){\line(1,0){2}}

\put(10,0){\line(0,1){4}}
\put(8,2){\line(1,0){4}}

\put(4.5,1){\circle*{0.11}}
\put(5.5,3.3){\circle*{0.11}}
\put(6.5,1.75){\circle*{0.11}}
\put(9.3,1.7){\circle*{0.11}}
\put(8.3,1.2){\circle*{0.11}}
\put(9.6,0.25){\circle*{0.11}}
\put(9.5,3.6){\circle*{0.11}}
\put(8.6,3.2){\circle*{0.11}}
\put(9.1,2.3){\circle*{0.11}}
\put(10.7,1.6){\circle*{0.11}}
\put(11.7,1.0){\circle*{0.11}}
\put(10.9,1.2){\circle*{0.11}}
\put(11.5,3.5){\circle*{0.11}}
\put(10.9,3.3){\circle*{0.11}}
\put(11.3,2.7){\circle*{0.11}}

\put(7.75,5.2){\circle*{0.11}}
\put(6.5,4.5){\circle*{0.11}}
\put(5.2,5.4){\circle*{0.11}}
\put(7.3,5.75){\circle*{0.11}}
\put(4.5,4.9){\circle*{0.11}}
\put(6.5,4.7){\circle*{0.11}}
\put(4.7,6.8){\circle*{0.11}}
\put(6.4,6.8){\circle*{0.11}}
\put(7.6,7.3){\circle*{0.11}}
\put(7.1,7.6){\circle*{0.11}}

\put(6.1,4.6){\circle*{0.11}}
\put(7.6,5.2){\circle*{0.11}}
\put(7.8,4.8){\circle*{0.11}}
\put(7.4,4.4){\circle*{0.11}}
\put(6.8,5.8){\circle*{0.11}}

\put(8.75,4.5){\circle*{0.11}}
\put(9.25,4.8){\circle*{0.11}}
\put(11.65,6.6){\circle*{0.11}}

\hskip 2cm
\setlength{\unitlength}{0.8cm}
\begin{picture}(10,10)(-6,1.5)

\put(0.80,3.5){${\bf x}_i$}
\put(7.25,1.6){$c$}
\put(7,4){\circle{0.15}}
\put(6.7,3.55){${\bf x}_c$}

\put(5,2){\line(0,1){4}}
\put(5,2){\line(1,0){4}}
\put(5,6){\line(1,0){4}}
\put(9,2){\line(0,1){4}}

\put(1,4){\circle*{0.10}}

\put(5.25,5.25){\circle*{0.10}}
\put(8.75,4){\circle*{0.10}}
\put(6.5,2.25){\circle*{0.10}}
\put(7,5.75){\circle*{0.10}}

\put(5.5,3){\circle*{0.10}}
\put(7.75,5){\circle*{0.10}}
\put(8,4.8){${\bf x}_j$}
\put(6,4.5){\circle*{0.10}}
\put(8,2.5){\circle*{0.10}}

\put(1,4){\line(1,0){5.9}}
\put(3.6,4.15){$R$}
\put(7.05,3.95){\line(1,-1){1.95}}
\put(8,3.2){$r_c$}
\end{picture}
\end{picture}
\caption{\footnotesize 
Details of treecode.
(a) tree structure of particle clusters. 
(b) particle-cluster interaction between particle ${\bf x}_i$ and cluster $c=\{{\bf x}_j\}$.
${\bf x}_c$: cluster center; $R$: particle-cluster distance; and $r_c$: cluster radius.}
\label{fig_treecode}
\end{figure}

We summarize the treecode algorithm
and
refer to previous work for more details~\cite{Barnes:1986,Duan:2001,Lindsay:2001,Li:2009}.
The matrix-vector product ${ A}{ x}$ for Eqs.~(\ref{eqbim_5})-(\ref{eqbim_6}) 
has the form of $n$-body potentials,
\begin{equation}
\label{particle-particle}
V_i = \sum_{{j=1}\atop{j \ne i}}^{N} q_jG({\bf x}_{i}, {\bf x}_{j}), \quad i=1,\ldots,N,
\end{equation}
where $G$ is a kernel,
${\bf x}_i, {\bf x}_j$ are centroids (also called particle locations in this context),
and
$q_j$ is a charge associated with ${\bf x}_j$. 
To this end, the $q_j$ in Eq.~(\ref{particle-particle}) is equivalent to the $\Delta s_j\phi_1({\bf x}_j)$ or $\Delta s_j\frac{\partial\phi_1({\bf x}_j)}{\partial{\nu}}$ in Eqs.~(\ref{eqbim_5})-(\ref{eqbim_6}) and $G$ is one of the kernels $K_{1-4}$.
To evaluate the potentials $V_i$ rapidly,
the particles ${\bf x}_i$ are divided into a hierarchy of clusters having a tree structure in a 2-D illustration as in
Fig.~\ref{fig_treecode}(a).
The root cluster is a cube containing all the particles
and
subsequent levels are obtained by dividing a parent cluster
into children~\cite{Barnes:1986}. 
The process continues until a cluster has fewer than $N_0$ particles ($N_0$ is a user-specified parameter representing the maximum number of particles per leaf, e.g.~$N_0=3$ in Fig.~\ref{fig_treecode}(a)).
Then $V_i$ is evaluated
as a sum of particle-particle interactions and particle-cluster interactions (depicted in Fig.~\ref{fig_treecode}(b)),
\begin{equation}
\label{particle_cluster_form}
V_i \,\approx
\sum_{c \, \in N_i}\sum_{{\bf x}_j \in c}q_jG({\bf x}_i,{\bf x}_j) \, +
\sum_{c \, \in F_i}\sum_{\|{\bf k}\|=0}^p a^{\bf k}({\bf x}_i,{\bf x}_c) \, m_c^{\bf k},
\end{equation}
where
$c$ denotes a cluster,
and
$N_i, F_i$ denote the near-field and far-field clusters of particle ${\bf x}_i$.
The first term on the right is a direct sum for particles ${\bf x}_j$ near ${\bf x}_i$,
and
the second term is a $p$th order Cartesian Taylor approximation about the cluster center ${\bf x}_c$
for clusters that are well-separated from ${\bf x}_i$~\cite{Li:2009}.
Cartesian multi-index notation is used with
${\bf k} = (k_1,k_2,k_3), k_i\,{\in}\,{\mathbb N}$ and $
\|{\bf k}\| = k_1+k_2+k_3$.
A particle ${\bf x}_i$ and a cluster $c$ are defined to be well-separated
if the multipole acceptance criterion (MAC) is satisfied,
$r_c/R \leq \theta$,
where
$r_c$ is the cluster radius,
$R=|{\bf x}_i-{\bf x}_c|$ 
is the particle-cluster distance 
and
$\theta$ is a user-specified parameter~\cite{Barnes:1986}.
\vskip 5pt

The accuracy and efficiency of the treecode is controlled 
by the combination of parameters including the order $p$, MAC parameter $\theta$, and maximum particles per leaf $N_0$.
Using the treecode,
the operation count for the matrix-vector product is $O(N\log N)$,
where $N$ is the number of particles ${\bf x}_i$,
and
the factor $\log N$ is the number of levels in the tree.

\subsubsection{Preconditioning}
\label{sec:preconditioning}

In order to precondition Krylov subspace methods in solving $Ax=b$, we design a scheme using left-preconditioning.  
Given a preconditioning matrix $M$, we consider the modified linear system $M^{-1}Ax=M^{-1}b$.  The solve then proceeds in two steps: (a) set $c = M^{-1}b$, and (b) solve $(M^{-1}A)x=c$ using GMRES.  
We therefore seek a preconditioner, $M$, such that two conditions are satisfied: \\
(1) $M$ is similar to $A$ such that $M^{-1}A$ has improved condition compared to $A$ and requires fewer GMRES iterations;\\
(2) $M^{-1}z=y$ can be efficiently computed, which is equivalent to solving $y$ from $My=z$.\\
Conditions (1) and (2) cannot be improved concurrently, thus a trade-off must be made. 


We design our preconditioner based on the observation that in the electrostatic interactions, 
which is also the interactions between boundary elements in solving integral equations, 
the short range interactions are smaller in number of interactions, but more significant in strength than the long range interactions, 
which are large in number of interactions and computationally more expensive. 
Due to their large number of interactions, the long interactions are calculated by multipole expansions. 
This offers the idea that for a preconditioner of $A$, we may construct $M$ to contain only short range interactions and to ignore long range interactions.  
To this end, we select short range interactions between elements on the same leaf only.
This choice of $M$ has great advantages in efficiency and accuracy for solving $My=z$.
As seen with details in \cite{Chen:2018}, by using a permutation operation, the $M$ matrix is a block diagonal matrix with $n$ blocks such that $M=\text{diag}\{M_1, M_2,\cdots,M_n\}$
thus $My=z$ can be solved using direct method e.g., LU factorization by solving each individual $M_iy_i=z_i$ for $i=1,\cdots,n$. Here each $M_i$ is a square nonsingular matrix, which represents the interaction between particles/elements on the $i$th leaf of the tree.  
It is worthy to note that the efficiency is not affected even when $M_i$ has a large condition number since a direct solver is used for solving $My=z$. Meanwhile, the computational cost for solving $My=z$ is $O(Nn_0^2)$ with $n_0$ the treecode parameter, maximum number of particles per leaf, as detailed in \cite{Chen:2018}.

\begin{figure}[htbp]
\begin{center}
\includegraphics[width=6in]{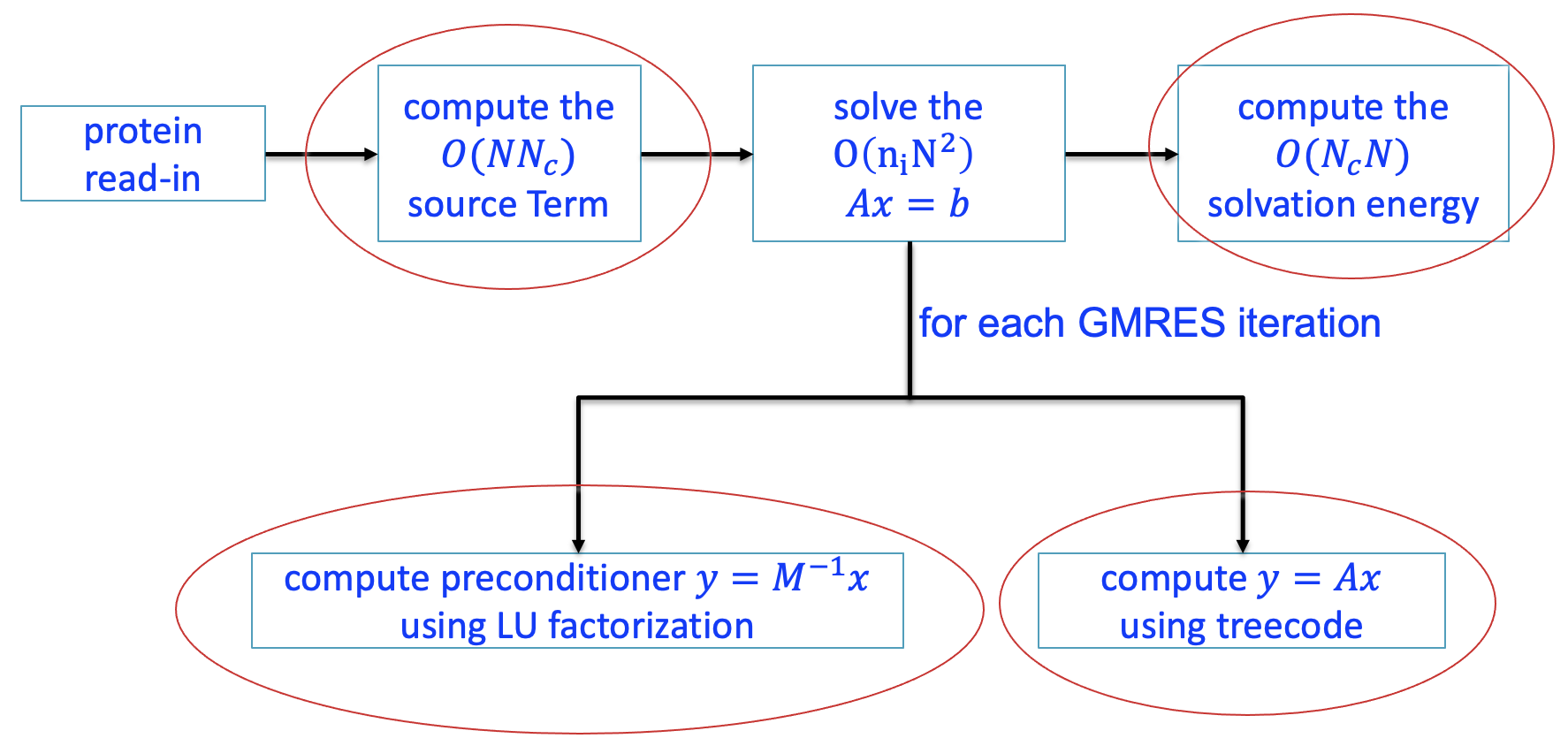}
\caption{pipeline for parallelized TABI solver}
\label{fig_MPIpipeline}
\end{center}
\end{figure}

\subsection{MPI-based Parallelization of the TABI Solver}
As illustrated by the red circled items in Fig.~\ref{fig_MPIpipeline}, our parallelization of the TABI solver focuses on the four stages of the pipeline: the $O(NN_c)$ source term, $O(n_iN\log N)$ matrix-vector product using treecode, the $O(n_iNn_0^2)$ preconditioner, and the $O(N_cN)$ solvation energy. Here, $N$ is the number of surface triangles, $n_0$ is the maximum number of particles per leaf, $N_c$ is number of partial charges, and $n_i$ is the number of GMRES iterations.  
Among these stages, the most time consuming and challenging component is the matrix-vector product using treecode. To this end, we investigate two possible strategies for computing $n$-body problems as in \cite{Chen:2021}, which are also briefly described below. In this work, we migrate these strategies to solving the boundary integral PB model. The numerical results in Section \ref{numericalresults} show high parallel efficiency from the optimized approach.

\begin{figure}[th!]
\begin{center}
\includegraphics[width=4.45in]{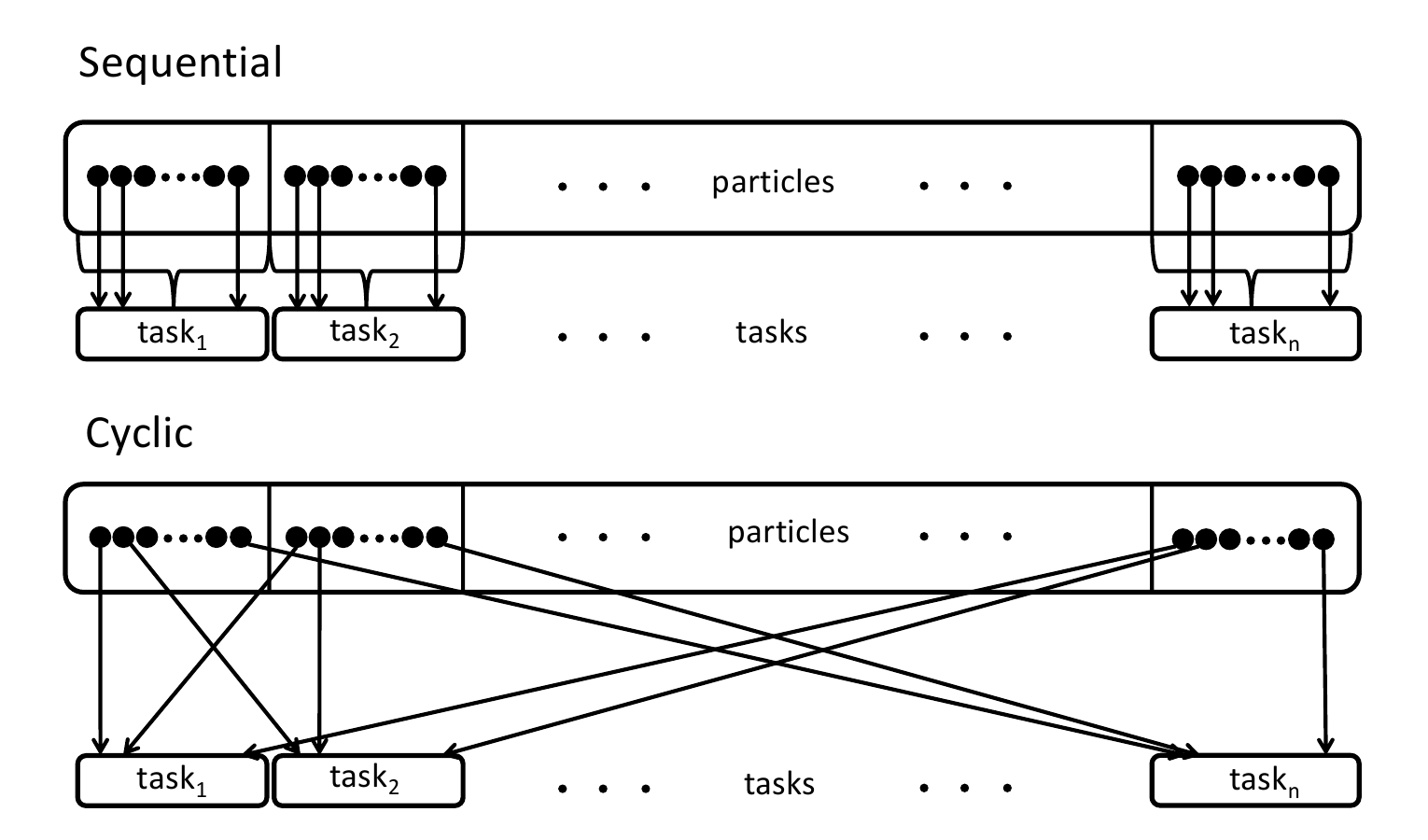}
\caption{methods for assigning target particles to tasks:
  sequential order (top) vs cyclic order (bottom)}
\label{fig_CycSeq}
\end{center}
\end{figure}

The initial and intuitive method to assign target particles to tasks
is to use {\it sequential ordering}, in which the 1st task handles
the first $N/n_p$ particles in a consecutive segment, the 2nd task
handles the next $N/n_p$ particles, etc. The illustration of this
job assignment is shown in the top of Fig.~\ref{fig_CycSeq}. However, when
examining the resulting CPU time on each task, we noticed starkly
different times on each task, indicating a severe load imbalance. This
may be understood by the fact that for particles at different
locations, the types of interactions with the other particles through the
tree can vary.  For example, a particle with only a few close neighbors
uses more particle-cluster interactions than particle-particle interactions,
thus requiring less CPU time than a particle with many close neighbors.
We also notice that for particles that are nearby one another,
their interactions with other particles, either by particle-particle
interaction or particle-cluster interaction, are quite similar, so
some consecutive segments ended up computing many more
particle-particle interactions than others that were instead dominated by
particle-cluster interactions.  Based on these observations, we
designed a {\it cyclic ordering} scheme to improve load balancing, as illustrated on the bottom of
Fig.~\ref{fig_CycSeq}.
In this scheme, particles nearby one another are uniformly distributed to different tasks. For example, for a group of particles close to each other, the first particle is handled by the first task, the second particle is handled by the second task, etc. The cycle repeats starting from the $(n_p+1)$-th particle.
The numerical results that follow demonstrate the significantly improved load balance from this simple scheme.

The pseudocode for our MPI-based parallel TABI solver using replicated data algorithm is given in Table~
\ref{pseudocode_MPI}. The identical trees are built on each task as in line 6. The four-stage MPI-based parallelization of the source term, matrix-vector product with treecode, preconditioning, and solvation energy occur in lines 7, 10, 12, and 17, respectively, followed by MPI communications.   

\begin{table}[htb]
\caption{
Pseudocode for MPI-based parallel TABI solver using replicated data algorithm.}
\begin{center}
\fbox{
\begin{tabular}{rl}
1 & on main processor \\
2 & \qquad read protein data \\
3 & \qquad call MSMS to generate triangulation \\
4 & \qquad copy protein data and triangulation to all other processors \\
5 & on each processor \\
6 & \qquad build local copy of tree \\
7 & \qquad compute assigned segment of source terms by direct sum \\
8 & \qquad\qquad copy result to all other processors \\
9 & \qquad set initial guess for GMRES iteration \\
10 & \qquad compute assigned segment of matrix-vector product $Ax$ by treecode \\
11 & \qquad\qquad copy result to all other processors \\
12 & \qquad compute assigned segment of solving $Mx=y$ for $x$ by LU factorization .\\
13 & \qquad\qquad copy result to all other processors \\
14 & \qquad test for GMRES convergence \\
15 & \qquad\qquad if no, go to step 10 for next iteration \\
16 & \qquad\qquad if yes, go to step 15 \\
17 & \qquad compute assigned segment of electrostatic solvation free energy by direct sum \\
18 & \qquad\qquad copy result to main processor \\
19 & on main processor \\
20 & \qquad add segments of electrostatic solvation free energy and output result \\
\end{tabular}
}
\end{center}
\label{pseudocode_MPI}
\vskip -30pt
\end{table}

\subsection{The GPU-accelerated DSBI Solver}
\begin{table}[!t]
\caption{
Pseudocode for DSBI-PB solver using GPU}
\begin{center}
\fbox{
\begin{tabular}{rl}
1 & On host (CPU)\\
2 & \qquad read biomolecule data (charge and structure) \\
3 & \qquad call MSMS to generate triangulation \\
4 & \qquad copy biomolecule data and triangulation to device \\
5 & On device (GPU)\\
6&  \qquad each thread concurrently computes and stores source terms for assigned triangles\\
7 & \qquad copy source terms on device to host\\
8 & On host \\
9 & \qquad set initial guess ${\bf x}_0$ for GMRES iteration and copy it to device \\
10 & On device\\
11 & \qquad each thread concurrently computes assigned segment of matrix-vector product ${\bf y}={\bf Ax}$ \\
12 & \qquad copy the computed matrix-vector ${\bf y}$ to host memory \\
~~13* & \qquad each thread concurrently solves its assigned portion of ${\bf Mx}={\bf y}$ \\
~~14* & \qquad copy the solution ${\bf x}$ to host memory \\
15 & On host\\
16 & \qquad test for GMRES convergence \\
17 & \qquad\qquad if no, generate new ${\bf x}$ and copy it to device, go to step 10 for the next iteration \\
18 & \qquad\qquad if yes, generate and copy the final solution to device and go to step 19 \\
19 & On device\\
20 & \qquad compute assigned segment of electrostatic solvation free energy \\
21 & \qquad copy computed electrostatic solvation free energy contributions to host \\
22 & On host\\
23 & \qquad add segments of electrostatic solvation free energy and output result \\

&* currently disabled\\
\end{tabular}
}
\end{center}
\label{pseudocode_GPU}
\end{table}

The pseudocode for the DSBI-PB solver using GPUs is given in Table~\ref{pseudocode_GPU}. 
%
In this psuedocode, we divide all the operations into those on host performed by the CPUs
and those on device performed by the GPUs. 
The three compute-intensive stages are computation of the source term, matrix-vector product, and solvation energy; each are computed on GPUs as shown in lines 6, 11, and 20, followed by a copy of the data from device to host. 
The host CPU takes care of all complicated and non-concurrent work. 
We note that lines 13 and 14 are still under investigation due to the considerations of parallel efficiency, and we disable these two lines in our current numerical implementation.  
One challenge is that the variance in sizes of the block matrices $M_i$ for $i=1,\ldots, n$ that compose the preconditioner $M$ lead to significant load imbalance on the GPU.  
The other is that the LU factorization used for solving the $M_iy_i=z_i$ is every inefficient on GPUs for our current implementation.
However, disabling the preconditioner within the GPU based parallelization could significantly increase computing time when $A$ is ill-conditioned.

\subsection{COVID-19 Proteins}
Coronaviruses are a persistent threat to global health. 
Viruses such as SARS in 2003, 
MERS in 2013, 
and the new SARS-CoV-2 in 2019 
emerge from animal populations and then infect humans. 
Coronaviruses contain a large genome 
which directs the synthesis of several dozen viral proteins. 
Structures of these proteins are used to better understand the diseases, and to 
develop new drugs and vaccines to fight coronaviruses.
In this work, we focus on proteins involved in the spreading and prevention of the COVID-19 virus. 
The virus genome in the form of an mRNA encodes proteins including replication/transcription complexes 
that make more RNA, structural proteins that construct new virions, 
and proteases (e.g.~61u7) that cut polyproteins into all of these functional pieces. 
The virus docks to target cells by binding the spike protein (e.g.~proteins 6crz, 6vxx, 6vsp, 6vsb) 
on the viral surface to its receptor, angiotensin-converting enzyme 2 (ACE2, e.g.~protein 6m17) 
on the target cell membrane. 
In addition, to test the infection of COVID-19 virus, 
we often identify its nucleocapsid proteins (e.g.~proteins 7act, 6yi3) 
by using antibodies (e.g.~proteins 7cr5, 7n3c, 7sts) that particularly bind to these nucleocapsids \cite{PDB}.   

In this work, we select a few COVID-19 related proteins and use our parallel PB solvers to calculate their electrostatic properties such as global solvation energy or local surface potential. 
These protein electrostatics can assist researchers in understanding a protein’s overall structure and function, their binding affinity to certain ligands, 
as well as their folding and enzyme catalysis characteristics. 

\begin{figure}[!tbp]
\begin{center}
\includegraphics[width=2.0in]{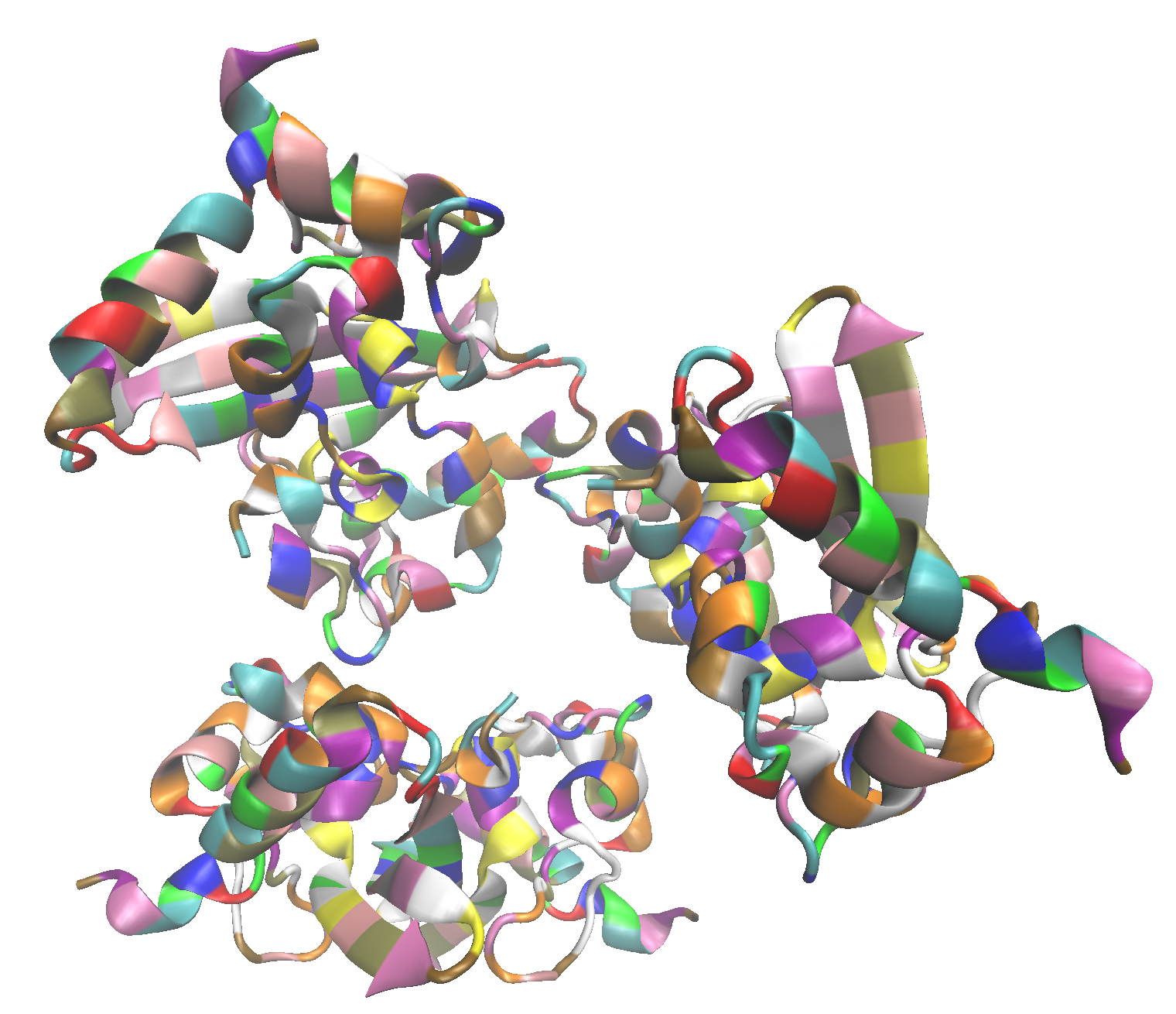}
\includegraphics[width=1.5in]{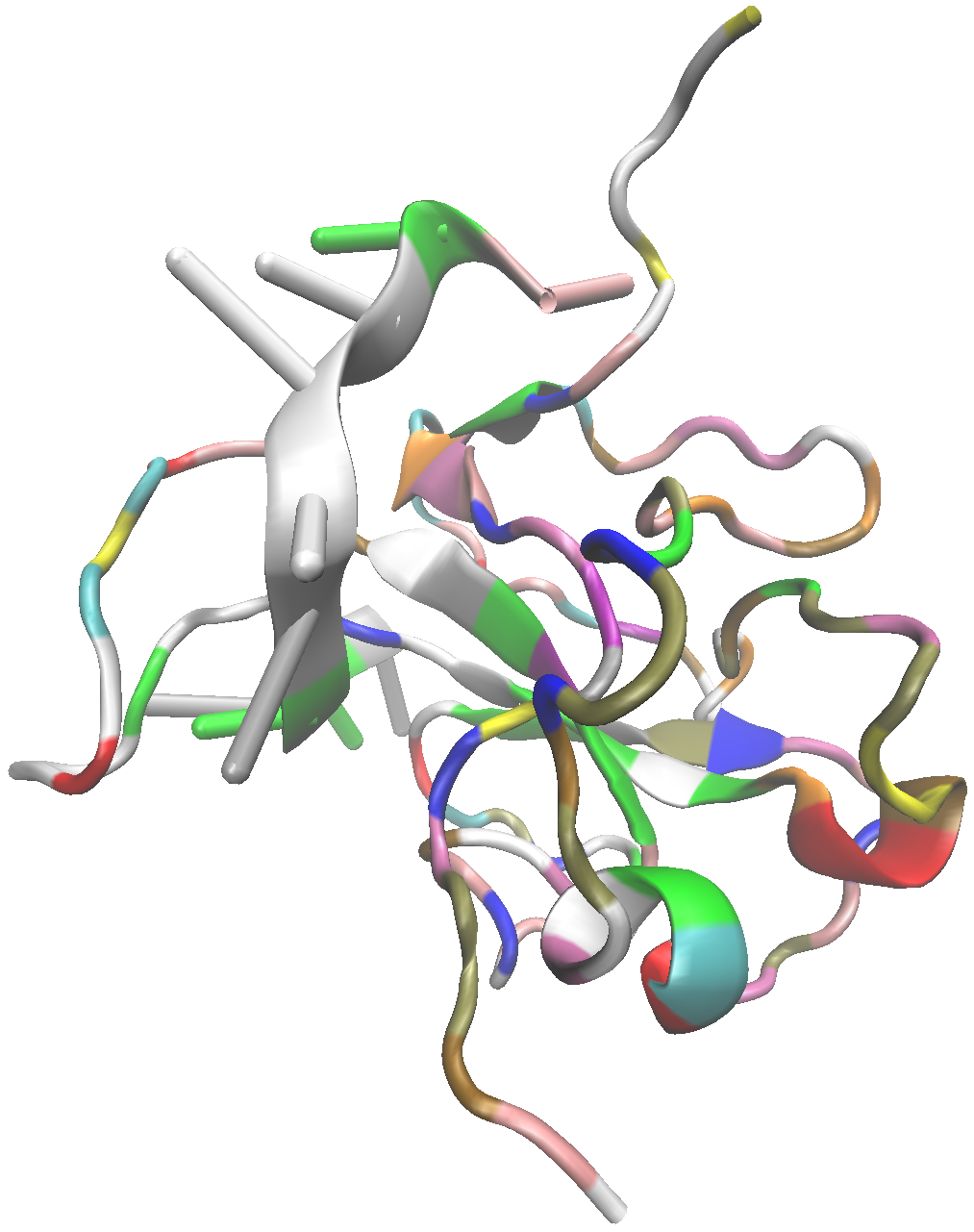}
\includegraphics[width=2.0in]{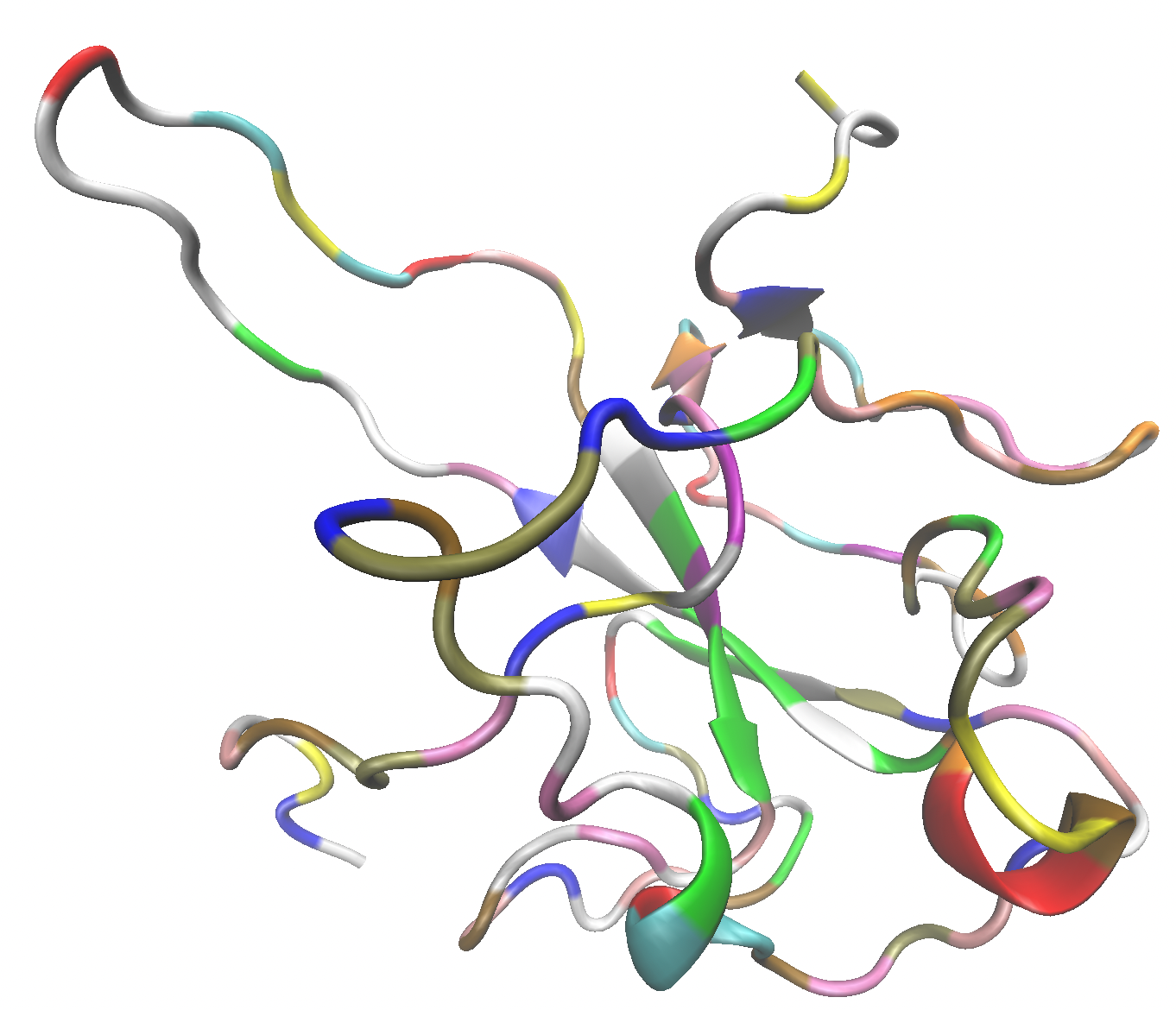}
\\
(a)\hskip 2in (b)\hskip 2in (c)\\
 \includegraphics[width=2.5in]{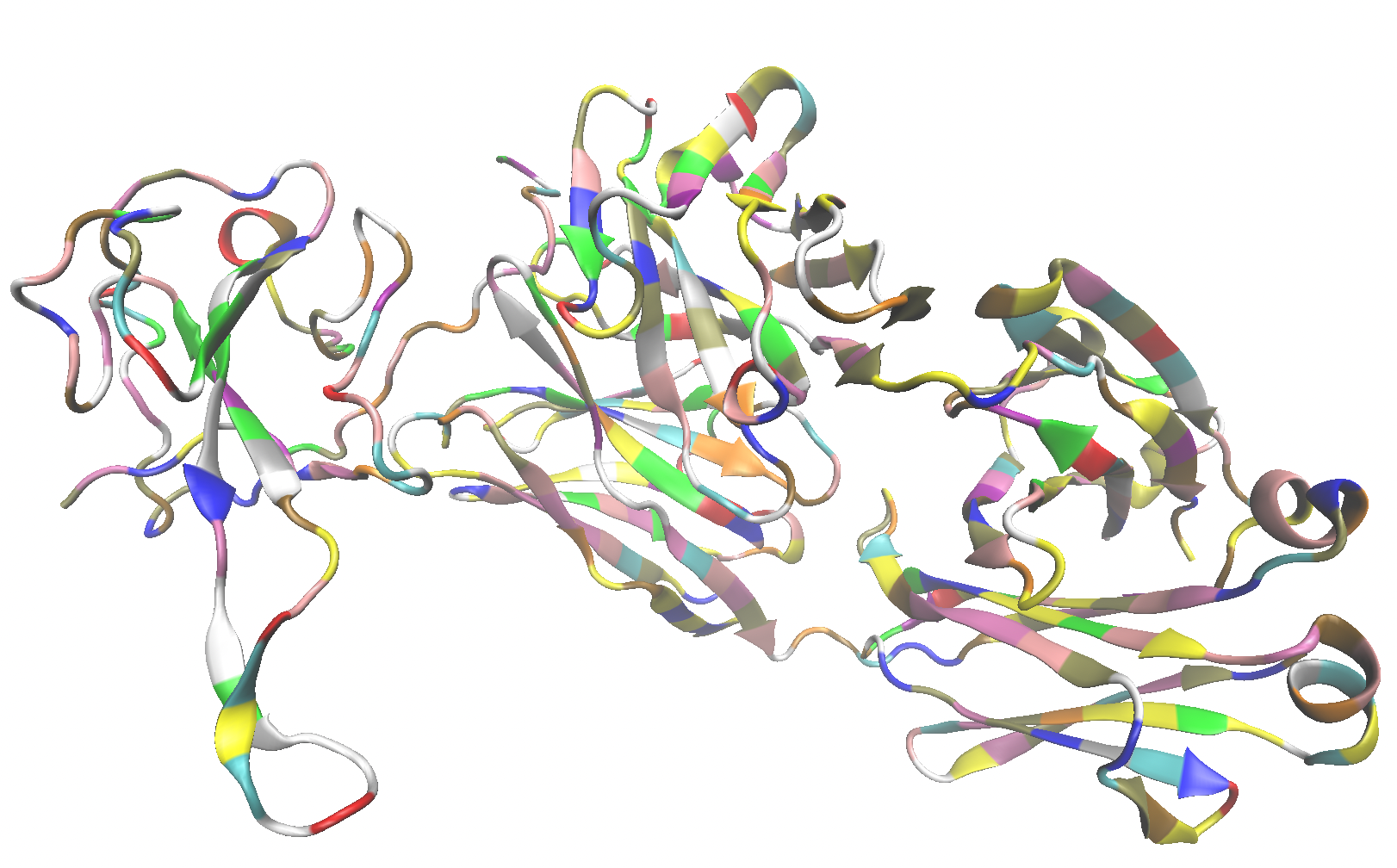}
 \includegraphics[width=1.0in]{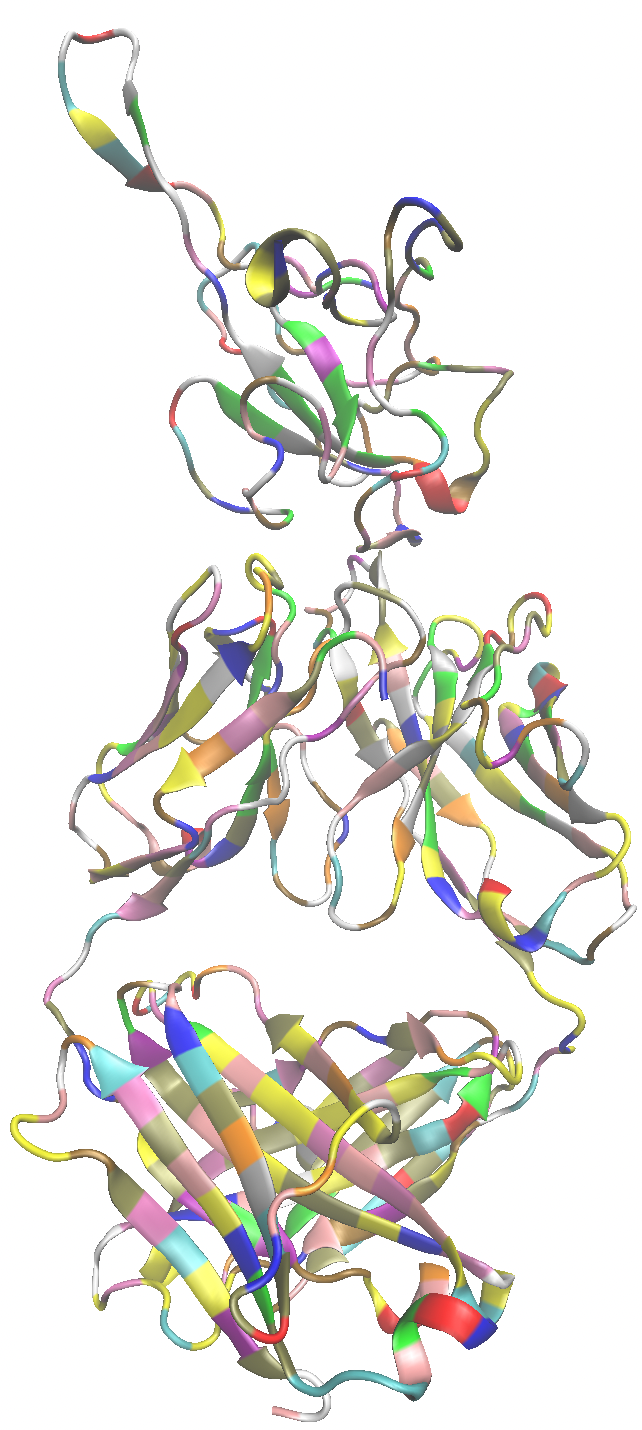}
 \includegraphics[width=2.5in]{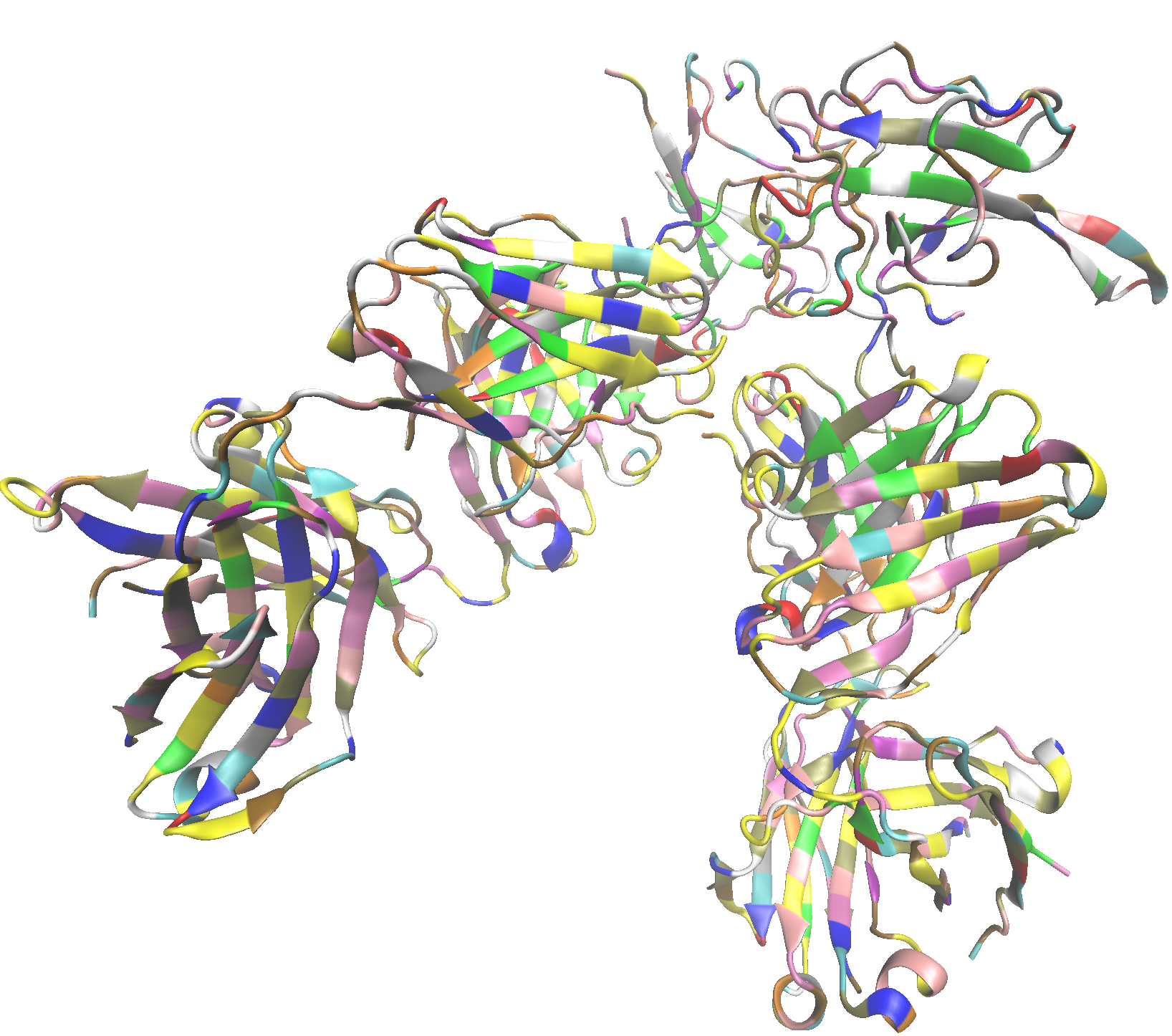}\\
(e)\hskip 2in (f)\hskip 2in (g)\\
\caption{COVID-19 related proteins used in our numerical simulations. (a) {\bf 6wji}: The C-terminal Dimerization Domain of Nucleocapsid Phosphoprotein from SARS-CoV-2; 
(b) {\bf 7act}: SARS-CoV-2 Nucleocapsid Phosphoprotein N-Terminal Domain in Complex with 10mer ssRNA; 
(c) {\bf 6yi3}: The N-terminal RNA-Binding Domain of the SARS-CoV-2 Nucleocapsid Phosphoprotein; 
(d) {\bf 7cr5}: Human Monoclonal Antibody with SARS-CoV-2 Nucleocapsid Protein NTD; 
(e){\bf 7n3c}: Human Fab S24-202 in the Complex with the N-Terminal Domain of Nucleocapsid Protein from SARS-CoV-2; 
(f) {\bf 7sts}: Human Fab S24-1379 in the Complex with the N-terminal Domain of Nucleocapsid Protein from SARS-CoV-2. \\}
\label{fig_Covid19Pros}
\end{center}
\end{figure}

In Fig.~\ref{fig_Covid19Pros}, we provide the cartoon structure of six COVID-19 related proteins. 
Protein 6wji \cite{Vandervaart:2023} in (a) is a dimerization domain, which is used to bring two nucleocapsids together. 
The connections of nucleocapsid dimers into bigger groups makes the viral structure that encases the RNA in the limited area within virus particles.
The SARS-CoV-2 nucleocapsid contains separate proteins which all perform different functions. 
A portion of the structure folds into an RNA-binding domain (protein 7act) \cite{Dinesh:2020} as shown in (b), featuring a groove that securely holds a brief segment of the viral genomic RNA. In contrast, the protein alone without the RNA-bound structure (protein 6yi3) is shown in (c) \cite{Dinesh:2020}. 
In COVID-19 prevention, home test kits for detecting SARS-CoV-2 infection 
rely on antibody proteins that specifically 
recognize nucleocapsids within a complex set of biomolecules in nasal samples. 
The antibodies they recognize differ based on the test-kit brand, which recognize different portions of the nucleocapsid, and we list a few here with protein 7cr5 \cite{Kang:2021} in (d), protein 7n3c in (e), and protein 7sts in (f). For these listed proteins, the PB model is solved using our parallel boundary integral PB solvers and numerical results are presented in the following section.

\section{Numerical Results}\label{numericalresults}

Our numerical results are generated on supercomputers sponsored by Southern Methodist University's Center for Research Computing (CRC).  
The MPI-based results are generated on M3 (\url{https://www.smu.edu/oit/services/m3}) 
and GPU-based results are generated on SuperPOD (\url{https://www.smu.edu/oit/services/superpod}). 

\subsection{Parallel Efficiency of MPI-based Computing}
We first check the parallel efficiency of our MPI-based algorithm with both sequential and cyclic schemes by computing the solvation energy on protein 7n3c at MSMS density of 12, which generates 529,911 boundary elements. 
We use up to 256 MPI tasks and Table~\ref{tb_MPI} shows the results. 
Column 1 shows the increasing numbers of MPI tasks.  Column 2 reports the total CPU time 
when the direct sum (DS)BI scheme is used to compute the electrostatic solvation free energy. 
Due to its $O(N^2)$ computational cost, the CPU time for the DSBI solver is overly long, even when 256 tasks are used. 
Columns 4 and 5 display the total CPU time and parallel efficiency for the TABI solver using the sequential and cyclic schemes, 
both of which are much faster compared to DSBI.   
Columns 8 and 9 focus more closely on the time required for a single matrix-vector product $Ax$, $\overline{t}_{Ax}$,
which we take as the average of the iteration’s maximum CPU time among all tasks,
\begin{equation}
\overline{t}_{Ax} = \frac{1}{n_i}\sum\limits_{k=1}^{n_i}\max\limits_{j} t_{Ax}^{j,k}
\label{eq_cpu}
\end{equation}
where $t_{Ax}^{j,k}$ is the CPU time to compute $Ax$ from the $j$th task in the $k$th GMRES iteration.

\begin{figure}[htbp]
\begin{center}
\includegraphics[width=3.2in]{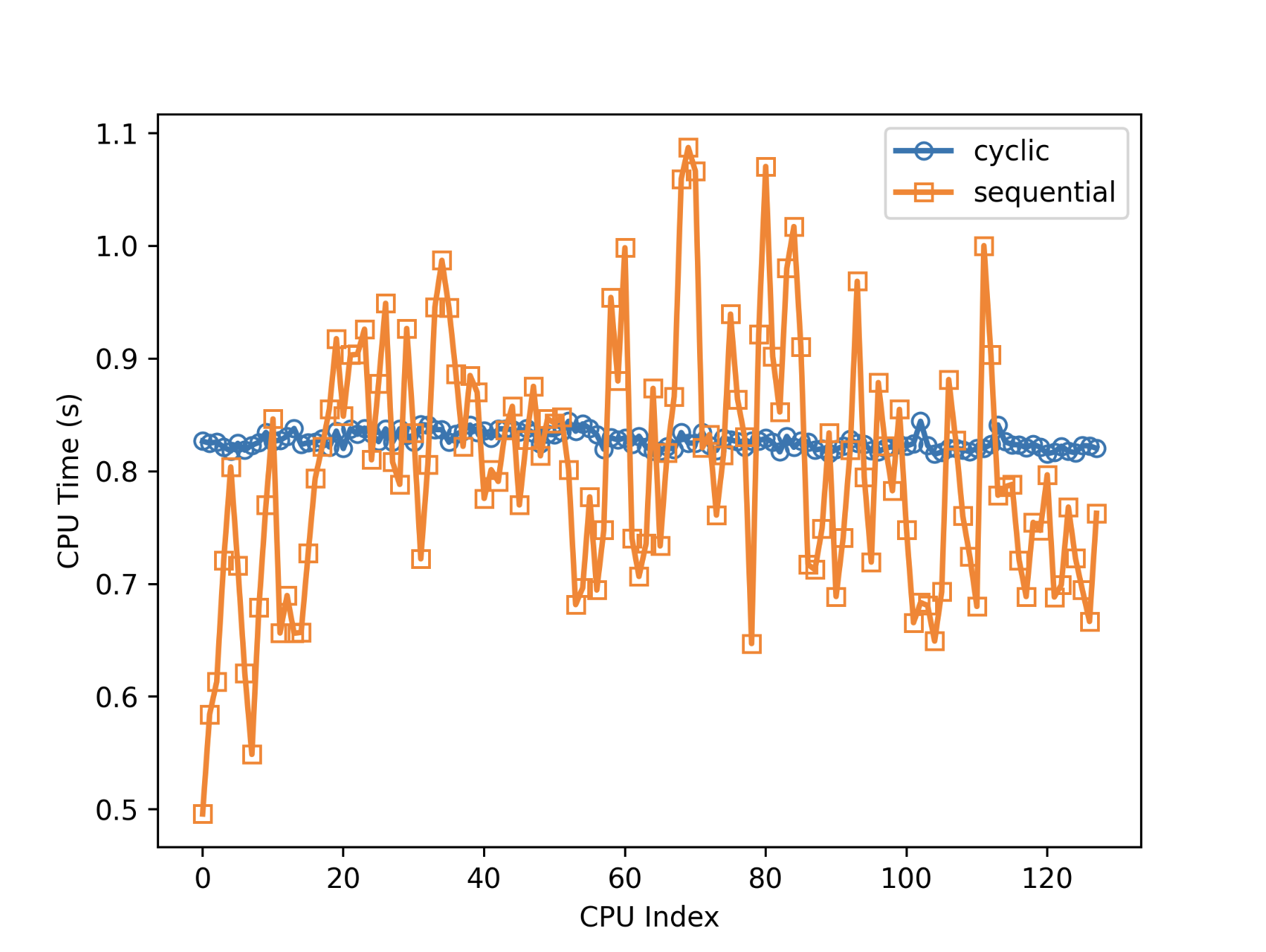}
\includegraphics[width=3.2in]{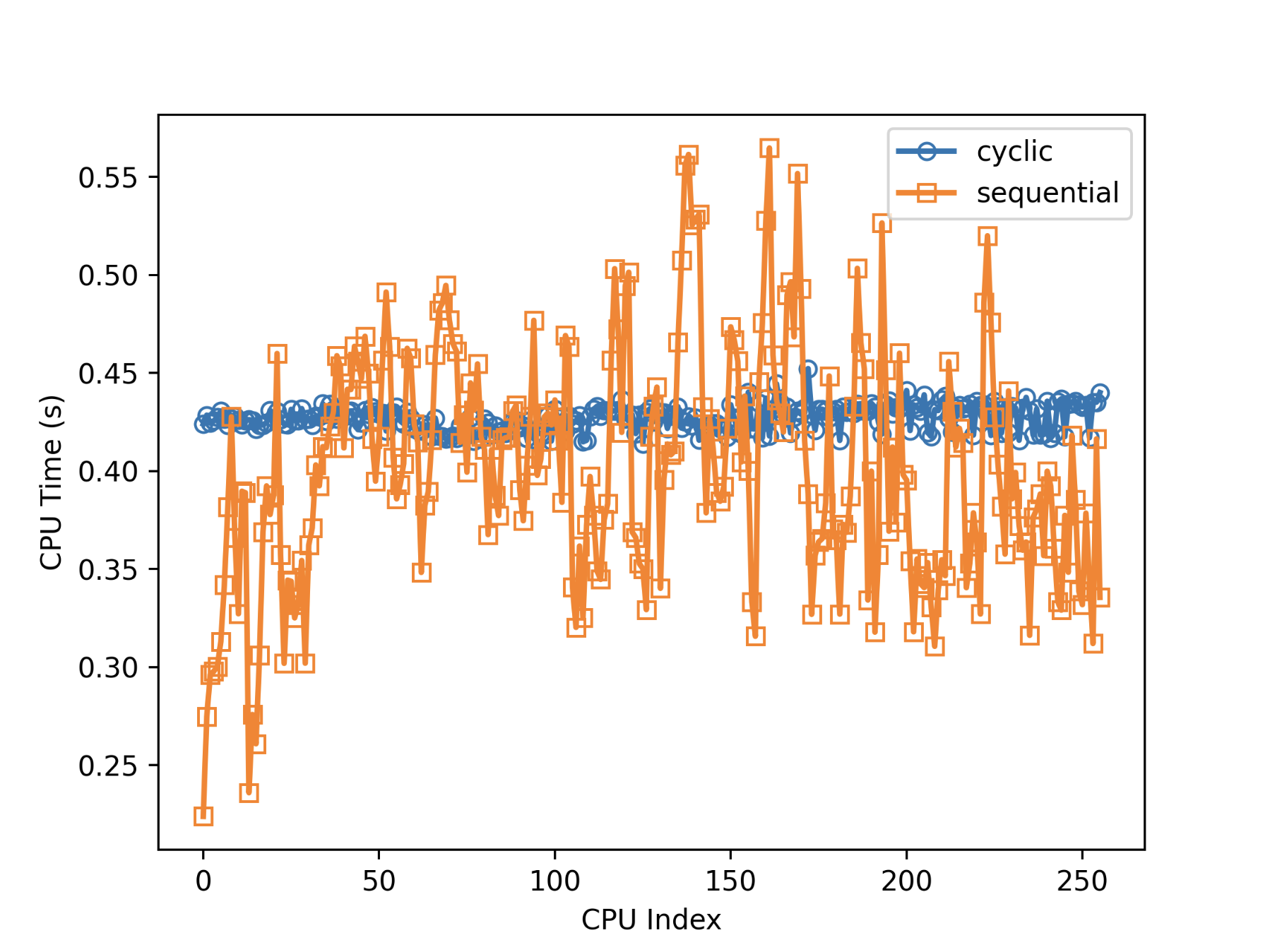}
\caption{MPI-based parallelization with sequential and cyclic schemes: left: 128 tasks, right: 256 tasks. The CPU time reported is $\overline{t}_{Ax}$, the averages GMRES iteration’s maximum CPU times among all tasks. 
}
\label{fig_MPI}
\end{center}
\end{figure}

\begin{table}[!htp]
{\small
\caption{\small CPU time and parallel efficiency (P.E.) for
  parallelized direct sum, sequentially parallelized treecode (seq.)
  and cyclically parallelized treecode (cyc.) for computing
  electrostatic solvation energy ({-6020.52} kcal/mol) for protein {7n3c}
  with 529,911 boundary elements. The treecode parameters are $\theta=0.8$, $N_0=100$,
  and $p=3$; The number of
  tasks $n_p$ ranges over $1,\ldots,256$. 
  The time for one $Ax$ $(\overline{t}_{Ax})$ is the average iteration's maximum CPU time over all tasks. 
  }
\begin{center}
\begin{tabular}{r||rr||rr|rr|rr|rr}
\hline\hline
$n_p$& \multicolumn{2}{|c||}{DSBI Solver} &  \multicolumn{8}{c}{TABI solver}\\\hline
&\multicolumn{2}{|c||}{Total Time}& \multicolumn{4}{|c|}{Total Time} & \multicolumn{4}{c}{Time for one $Ax$ ($\overline{t}_{Ax}$,)}  \\\hline
& CPU (s) & P.E. (\%) & \multicolumn{2}{|c|}{CPU (s) } & \multicolumn{2}{c}{P.E. (\%)} & \multicolumn{2}{|c|}{CPU (s) } & \multicolumn{2}{c}{P.E. (\%)}\\\hline
& & & seq. & cyc. & seq. & cyc. & seq. & cyc. & seq. & cyc.\\\hline
1 & 106063.17 & 100.00  & 1874.88 & 1873.60 & 100.00  & 100.00 & 89.75 & 89.60 & 100.00 & 100.00\\
2 & 53132.86  & 99.81  & 971.25  & 967.12  & 96.52  & 96.87  & 45.49 & 45.22 & 98.63 & 99.07  \\
4 & 26549.87  & 99.87  & 561.25  & 502.60  & 83.51  & 93.20  & 25.69 & 22.57 & 87.34 & 99.26  \\
8 & 13291.47  & 99.75  & 321.42  & 285.25  & 72.91  & 82.10  & 13.94 & 12.02 & 80.46 & 93.22  \\
16 & 6710.06  & 98.79  & 171.41  & 158.04  & 68.36  & 74.09  & 6.43  & 5.77  & 87.30 & 97.08  \\
32 & 3928.71  & 84.37   & 128.13  & 114.73  & 45.73  & 51.03  & 3.99  & 3.26  & 70.22 & 85.84  \\
64 & 2022.84  & 81.93   & 99.75  & 90.81  & 29.37  & 32.24  & 2.14  & 1.66  & 65.55 & 84.24  \\
128 & 1042.49 & 79.48   & 79.82   & 76.83   & 18.35  & 19.05  & 1.08  & 0.85  & 64.65 & 82.77  \\
256 & 554.50 & 74.72   & 71.70   & 71.16   & 10.21   & 10.28   & 0.56  & 0.45  & 62.27 & 78.61  \\
\hline\hline
\end{tabular}
\end{center}
\label{tb_MPI}
}
\end{table}
Parallel efficiencies are displayed in Columns 3, 6, 7, 10 and 11. 
The parallelization of the DSBI solver shows high efficiency as seen in column 3. 
This is due to the simplicity of the algorithm. 
Other than the four stages identified in Figure \ref{fig_MPIpipeline}, there is very little serial 
computation or communication required.  However, the parallel efficiency of the TABI solver is not as good as the DSBI solver, as shown in Columns 6 and 7. This is primarily due to the use of treecode, which has some serial time for building the tree and computing the moments. The serial time is relatively short when $n_p$ is small but becomes increasingly significant as $n_p$ grows and the time spent within  parallelized stages decreases. 
If we focus specifically on the parallelization of the treecode in computing $Ax$, Columns 10 and 11 show a high degree of parallel efficiency. 
We can also observe that the cyclic scheme significantly improves the parallel efficiency in comparison with the sequential scheme. 
However, due to the very small fraction of runtime spent in computing matrix-vector products as $n_p$ increases, the overall parallel efficiencies from Columns 6 and 7 do not show a significant difference between the sequential and cyclic schemes.  To more carefully examine the performance differences between the sequential and cyclic schemes, in Fig.~\ref{fig_MPI} we plot $\overline{t}_{Ax}$ from Eq.~\eqref{eq_cpu} when 128 and 256 MPI tasks are used. 
It is evident that the cyclic scheme has reduced variance 
compared with the sequential scheme owing to its advantage in load balance.

\subsection{MPI-based TABI solver {\it vs} GPU-accelerated DSBI solver}

Next, we compute the solvation energy for the six COVID-19 proteins introduced previously using MSMS with density equal to 12 to provide sufficient detail of the molecular surface. We use both the MPI-based TABI solver and the GPU-accelerated DSBI solver. For a reasonable computing power comparison, we use 64 CPU cores for the MPI-related computing and 1 GPU card for the GPU-related computing. Table~\ref{tb_proteins} shows the simulation results.
Column 1 is the PDB ID for proteins in ascending sequence of their size followed by the number of atoms in column 2, number of boundary elements in column 3, and the areas of the solvent excluded surface in column 4. Columns 5 and 6 are the number of GMRES iterations, from which we can see that the TABI solver has much improved condition number in comparison with the DSBI solver thanks to the TABI preconditioner from Section \ref{sec:preconditioning}. The solvation energies are reported in columns 7 and 8, which are sufficiently close. The differences are caused by the treecode approximation, the preconditioning scheme, and the error tolerance achieved when the iteration is stopped. 
Note for calculating protein electrostatic solvation energy, we don't have an exact value to compare. If the DSBI solver converges before reaching the maximum number of allowed GMRES iterations, its result should be more accurate than that from TABI solver since Treecode and Preconditioner could add extra approximations. For example, the $E_\text{sol}^\text{GPU}$ results from proteins 6yi3, 7act, 7n3c, 6wji should be more accurate than the  $E_\text{sol}^\text{MPI}$ results for these proteins. 
However, if the GMRES allowed maximum number of iteration has been reached for examples for proteins 7cr5 and 7sts, 
on which the DSBI solver stopped when 100 iterations are reached 
while the accuracy did not meet the $10^{-4}$ threshold, 
we can not say for sure whether $E_\text{sol}^\text{GPU}$ result or $E_\text{sol}^\text{MPI}$ result is more accurate. 
The computation times are shown in columns 9 and 10, which demonstrates that the computing power between 64 CPUs and 1 GPU are comparable. However, the algorithms (with preconditioning vs without preconditioning, direct sum vs treecode) can make a substantial difference for ill-conditioned or larger systems.  For example, for protein 7sts, with nearly one million boundary elements, the MPI-based TABI solver is significantly faster than the GPU-based DSBI because of the ill-conditioned system, the large size of the problem, and the use of treecode or not.   

\begin{table}[htp]
\caption{Computing electrostatic solvation energies in (kcal/mol) for the involved proteins: ionic strength = 0.15M; $\epsilon_1=1$, $\epsilon_2=80$; MSMS \cite{Sanner:1996} 
density=12; $N_c$ is the number of atoms/charges, $N$ is the number of boundary elements, $n_i$ is the number of GMRES iterations, $S_\text{ses}$ is the solvent excluded surface area, and $E_\text{sol}$ is the electrostatic solvation energy. 
}
{\small		
  \begin{center}
		\begin{tabular}{c|c|c|r|r|r|r|r|r|r}
		\hline
		PDB & $N_c$ & $N$ & $S_\text{SES}$ & $n_i^{\text{MPI}}$ & $n_i^{\text{GPU}}$   & $E_\text{sol}^\text{MPI}$ & $E_\text{sol}^\text{GPU}$ & $t^\text{MPI}$ (s)  & $t^\text{GPU}$ (s) \\
		\hline
		6yi3 & 2083  & 169,968  & 7516.44  & 10 & 10   & -1941.81  & -1945.18  & 14.76  & 8.96   \\
		7act & 2352  & 188,054  & 8286.70  & 14 & 14   & -1893.88  & -1934.49  & 21.35  & 17.39  \\
		7cr5 & 8133  & 513,226  & 22524.23 & 16 & 100+ & -5713.52  & -5786.69  & 89.52  & 695.17 \\
		7n3c & 8459  & 530,084  & 23244.85 & 19 & 17   &  -6020.52  & -6013.68  & 99.13 & 132.20 \\
		6wji & 10182 & 641,266  & 28116.88 & 13 & 14   & -14009.55 & -14016.02 & 112.82 & 152.71 \\ 
		7sts & 15797 & 993,572  & 43457.63 & 26 & 100+ & -11622.63 & -11583.26 & 422.67 & 2544.70 \\
		\hline		
		\end{tabular}
		\end{center}
  }
\label{tb_proteins}
\end{table}%

We then further investigate under what conditions we should choose betwen using the GPU-accelerated DSBI solver or MPI-based TABI solver. The following example, whose result is shown in Table~\ref{tb_6yi3}, gives some important guidance. In this example, we compute the solvation energy for protein 6yi3 for increasing values of the MSMS density ($d$), giving rise to increased problem sizes, as shown in columns 1 and 2. Columns 3 and 4 show the similar solvation energy computed with these two approaches.  
Columns 5 and 6 report the number of GMRES iterations. From these close results, we see that the discretized system for this protein is well conditioned thus the preconditioning scheme has limited effect. We solve the problem using 1 CPU core and report the time in column 7 for reference. Then we report the time for solving the problem using 64 MPI tasks and one A100 GPU card in columns 8 and 9. The result indicates that for a protein whose discretized system is well-conditioned, when the number of boundary elements is less than 250,000, we should use the GPU-accelerated DSBI solver, since the smaller the system the better the GPU-accelerated DSBI solver compares against the MPI-based TABI solver. 
If the conditioning of $A$ shows a pressing need for preconditioning, the threshold number will be smaller for the GPU-accelerated DSBI solver.
The rapid GPU performance at least gives us the hope to perform molecular dynamics or Monte Carlo simulation for small and middle-sized proteins using GPUs. For example, if 50,000 boundary elements can reasonably describe the given protein, a single PB equation solution only takes about one second using one GPU card, in comparison with 4 seconds on a 64-core cluster.

\begin{table}[htp]
\caption{Computing electrostatic solvation energies in (kcal/mol)  for the protein 6yi3 at different MSMS densities: ionic strength = 0.15M; $\epsilon_1=1$, $\epsilon_2=80$; $d$ is the MSMS density, $N$ is the number of boundary elements, $n_i$ is the number of GMRES iterations, $E_\text{sol}$ is the electrostatic solvation energy. Results are generated using KOKKOS and MPI on ManeFrame III; MPI results are from using 64 tasks; GPU results are from using one A100 GPU. 
}
\label{tb_6yi3}
{\small		
  \begin{center}
		\begin{tabular}{c|r|r|r|r|r|r|r|r}
		\hline
		$d$ & $N$~~~ & $E_\text{sol}^\text{MPI}$& $E_\text{sol}^\text{GPU}$ & $n_i^\text{CPU}$ & $n_i^\text{GPU}$ &  $t_\text{CPU}$ (s)  & $t_\text{MPI }$ (s) & $t_\text{GPU}$ (s) \\
		\hline
		2 &28,767&-2057.61&-2056.26&10&10&29.34&2.76&0.83 \\
		4 &56,127&-1999.01&-1997.03&10&10&66.40&4.46&1.52\\
		6 &84,903&-1968.00&-1966.87&10&16&108.52&7.05&4.98 \\
		8 &110,307&-1954.62&-1952.23&10&10&145.13&9.35&4.32\\
		12 &169,955&-1945.18&-1941.81&10&10&240.54&14.76&8.91 \\
		16 &229,901&-1940.96&-1936.87&10&11&340.82&19.86&17.66 \\
        18 &257,236&-1938.27&-1933.67&10&11&385.96&{\bf 21.69}& {\bf 23.73}\\
        20 &287,202 &-1937.18&-1931.77&10&12 &438.86&24.54&28.73\\
		24 &343,806 &-1933.63&-1928.65&10&11 &534.31&35.13&38.62\\
        28 &407,196&-1933.04&-1927.56&10&12&653.06&41.84&55.18\\
        32.5 &471,307&-1931.76&-1926.04&10&12&760.12&51.23&77.83 \\
		64 &946,335&-1928.51&-1921.81&10&13&1701.06&145.65&311.07 \\
		\hline		
		\end{tabular}
		\end{center}
  }
\end{table}%

\begin{figure}[htbp]
\includegraphics[width=2.3in]{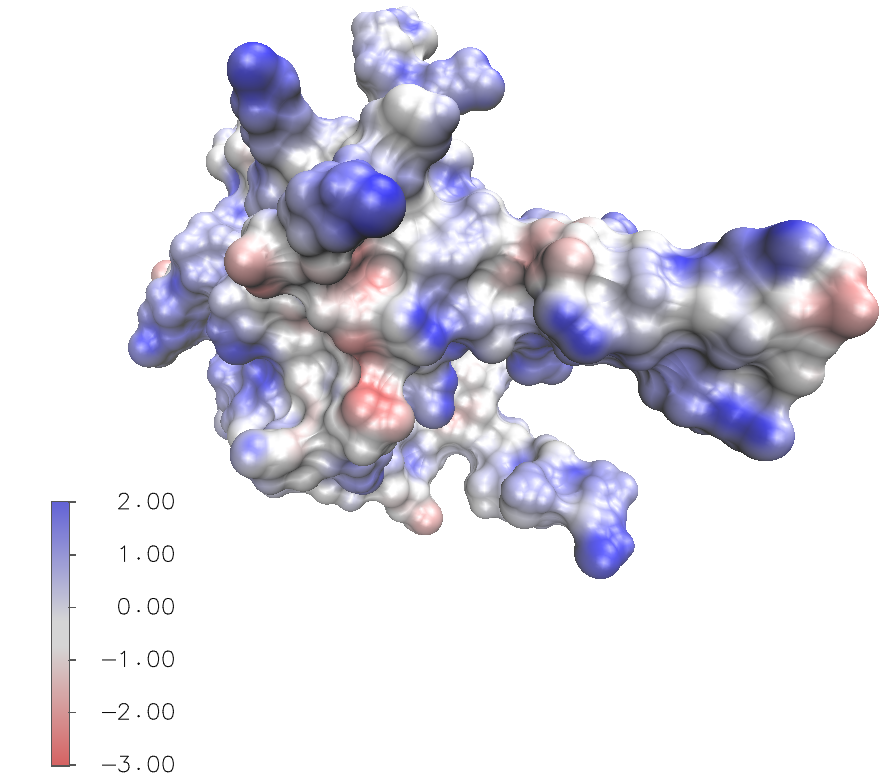}
 \includegraphics[width=2.0in]{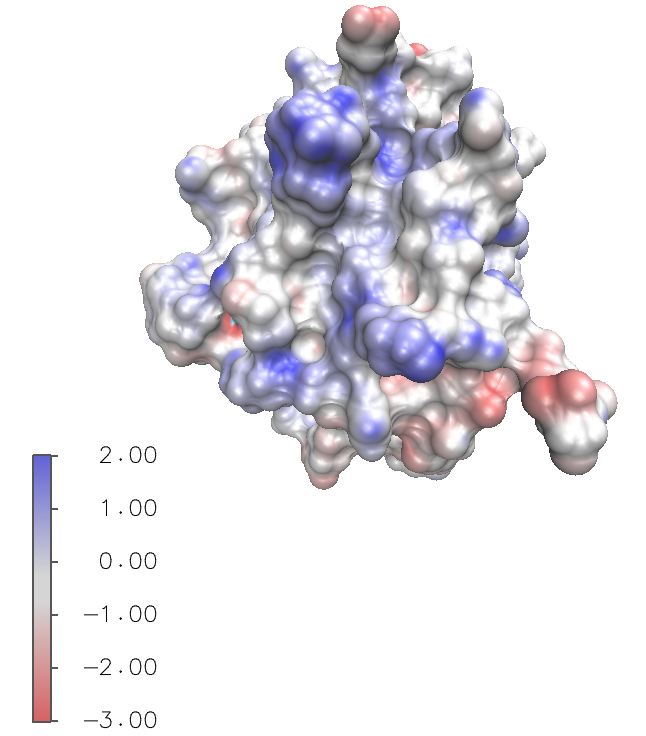}
 \includegraphics[width=2.3in]{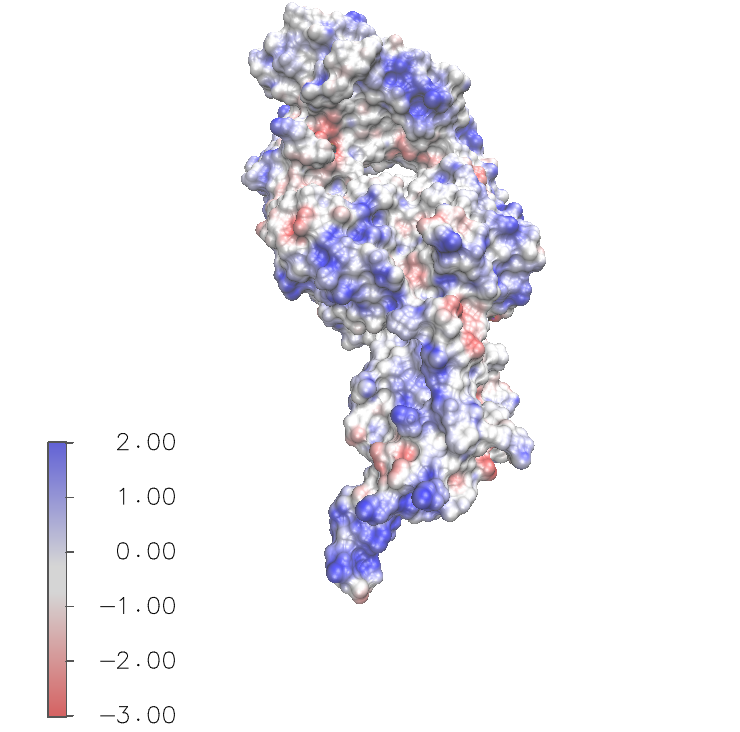}\\
\caption{Color coded electrostatic surface potential in kcal/mol/$e_c$ on the molecular surface of proteins 6yi3 (left), 7act (middle), and 7n3c (right); plot is drawn with VMD \cite{Humphrey:1996}.}
\label{fig_pot}
\end{figure}

We note that the solution to the boundary integral PB equation gives both the electrostatic potential and its normal derivative on the molecular surface. We can plot the potential on the surface elements. The color-coded potential can provide guidance on the docking site for the ligand, or offer other insights pertaining to protein-protein interactions. Some examples of this kind of visualization are shown in Fig.~\ref{fig_pot}.

\section {Software Dissemination}
The source code for all solvers and examples used in this project are available on GitHub. The MPI code can be found on \url{https://github.com/elyssasliheet/tabi_mpi_code} maintained by SMU graduate student Elyssa Sliheet. The GPU code can be found on \url{https://github.com/yangxinsharon/bimpb-parallelization} maintained by SMU graduate student Xin Yang.

\section {Concluding Remarks}
In this project, we investigate the practical application of the PB model on selected proteins which play significant roles in the spread, treatment, and prevention of COVID-19 virus diseases. 
To this end, we solved the boundary integral form of the PB equation on the molecular surfaces of these proteins.  These calculations produce both the electrostatic solvation free energy as a global measurement and the electrostatic surface potential for local details of the selected proteins. 
We investigated the parallel performance of two competing solvers for the boundary integral PB equations on these selected proteins. 
By considering the advantages of current algorithms and computer hardware, we focused on the parallelization of the TABI solver using MPI on CPUs and the DSBI solver using KOKKOS on GPUs. Our numerical simulations show that the DSBI solver on one A-100 GPU is faster than the TABI solver with MPI on 64 CPUs when the number of elements is smaller than 250,000. 
When both GPU and MPI are available and the triangulation quality is good enough so that the TABI preconditioner is not needed for GMRES convergence, we recommend that the GPU-accelerated DSBI PS solver be used when the number of boundary elements is below 250,000.  Otherwise, the MPI-based TABI should be used.  If the number of elements becomes so large such that the memory on a CPU task cannot hold an entire tree, we recommend consideration of a domain-decomposition MPI scheme \cite{Vaughn:2020a,Chen:2021a,Wilson:2021,Salmon:1986}.  We note that the memory usage for TABI scales linearly with problem size. When one million boundary elements are used, the memory usage is a little bit over 1GB. Thus for popular tasks on clusters with at least 64G memory per MPI rank, we can handle problems as large as approximately 64 million boundary elements, which is sufficient for simulating middle-large proteins with up to tens of thousands atoms. For even larger biomolecules, e.g. the viral capsids of Zika or H1N1 virus with up to tens of millions atoms \cite{Wilson:2022a}, domain decomposition approach can be considered \cite{Wilson:2021}. \\ 

\noindent
{\bf Acknowledgments} 

This work of XY, ES, and WG was supported in part by the National Science Foundation (NSF) grants DMS-2110922 and DMS-2110869. 
ES was also partially support by the NSF RTG-1840260 grant. 
RI was also supported in part by SMU's Hamilton Scholar and Undergraduate Research Assistantships (URA) programs. 

We thank the SMU Mathematics Department for providing the parallel computing class MATH 6370, which systematically trains graduate students on parallelization strategies, schemes, and experience. We also thank the SMU Center for Research Computing (CRC) for proving computing hardware. These resources combined to make this project possible.\\ 

\noindent 
{\bf Data Availability} \\
Enquiries about data availability should be directed to the authors. \\

\noindent
{\bf \large Declarations}\\
{\bf Conflict of interest} \\
The authors have not disclosed any competing interests.


\bibliographystyle{spmpsci}
\bibliography{pb_review_um}

\begin{thebibliography}{10}
\providecommand{\url}[1]{{#1}}
\providecommand{\urlprefix}{URL }
\expandafter\ifx\csname urlstyle\endcsname\relax
  \providecommand{\doi}[1]{DOI~\discretionary{}{}{}#1}\else
  \providecommand{\doi}{DOI~\discretionary{}{}{}\begingroup
  \urlstyle{rm}\Url}\fi

\bibitem{PDB}
{\url{http://www.rcsb.org/pdb/home/home.do}}

\bibitem{Barnes:1986}
Barnes, J., Hut, P.: A hierarchical {O(NlogN)} force-calculation algorithm.
\newblock Nature \textbf{324}, 446--449 (1986).
\newblock \urlprefix\url{http://dx.doi.org/10.1038/324446a0}

\bibitem{Beard:2001}
Beard, D.A., Schlick, T.: Modeling salt-mediated electrostatics of
  macromolecules: the discrete surface charge optimization algorithm and its
  application to the nucleosome.
\newblock Biopolymers \textbf{58}, 106--115 (2001)

\bibitem{Bedorf:2012}
B{\'e}dorf, J., Gaburov, E., {Portegies Zwart}, S.: {A sparse octree
  gravitational N-body code that runs entirely on the GPU processor}.
\newblock Journal of Computational Physics \textbf{231}(7), 2825--2839 (2012)

\bibitem{Belleman:2008}
Belleman, R.G., B{\'e}dorf, J., {Portegies Zwart}, S.F.: {High performance
  direct gravitational N-body simulations on graphics processing units II: An
  implementation in CUDA}.
\newblock New Astronomy \textbf{13}(2), 103--112 (2008)

\bibitem{Burtscher:2011}
Burtscher, M., Pingali, K.: An efficient {CUDA} implementation of the
  tree-based Barnes-Hut {N}-body algorithm, pp. 75--92.
\newblock Elsevier Inc. (2011).
\newblock \doi{10.1016/B978-0-12-384988-5.00006-1}

\bibitem{Callenberg:2010}
Callenberg, K.M., Choudhary, O.P., de~Forest, G.L., Gohara, D.W., Baker, N.A.,
  Grabe, M.: Apbsmem: A graphical interface for electrostatic calculations at
  the membrane.
\newblock PLoS ONE \textbf{5}(9), 1--12 (2010).
\newblock \doi{10.1371/journal.pone.0012722}.
\newblock \urlprefix\url{http://dx.doi.org/10.1371%2Fjournal.pone.0012722}

\bibitem{Chen:2018}
Chen, J., Geng, W.: {On preconditioning the treecode-accelerated boundary
  integral (TABI) Poisson-Boltzmann solver}.
\newblock J. Comput. Phys. \textbf{373}, 750--762 (2018)

\bibitem{Chen:2021a}
Chen, J., Geng, W., Reynolds, D.: Cyclically paralleled treecode for fast
  computing electrostatic interactions on molecular surfaces.
\newblock Comput. Phys. Commun. \textbf{260}, 107742 (2021)

\bibitem{Chen:2021}
Chen, J., Geng, W., Wei, G.W.: {MLIMC: Machine learning-based implicit-solvent
  Monte Carlo}.
\newblock Chinese Journal of Chemical Physics \textbf{34}(6), 683--694 (2021).
\newblock \doi{10.1063/1674-0068/cjcp2109150}.
\newblock \urlprefix\url{https://doi.org/10.1063/1674-0068/cjcp2109150}

\bibitem{Chen:2023}
Chen, J., Tausch, J., Geng, W.: {A Cartesian FMM-accelerated Galerkin boundary
  integral Poisson-Boltzmann solver}.
\newblock Journal of Computational Physics \textbf{478}, 111981 (2023)

\bibitem{Cherezov:2007}
Cherezov, V., Rosenbaum, D.M., Hanson, M.A., Rasmussen, S.G.F., Thian, F.S.,
  Kobilka, T.S., Choi, H.J., Kuhn, P., Weis, W.I., Kobilka, B.K., Stevens,
  R.C.: {High-Resolution Crystal Structure of an Engineered Human
  beta2-Adrenergic G Protein{\textendash}Coupled Receptor}.
\newblock Science \textbf{318}(5854), 1258--1265 (2007).
\newblock \doi{10.1126/science.1150577}.
\newblock \urlprefix\url{http://science.sciencemag.org/content/318/5854/1258}

\bibitem{Dinesh:2020}
Dinesh, D.C., Chalupska, D., Silhan, J., Koutna, E., Nencka, R., Veverka, V.,
  Boura, E.: {Structural basis of RNA recognition by the SARS-CoV-2
  nucleocapsid phosphoprotein}.
\newblock PLOS Pathogens \textbf{16}(12), 1--16 (2020)

\bibitem{Dong:2003}
Dong, F., Vijaykumar, M., Zhou, H.X.: Comparison of calculation and experiment
  implicates significant electrostatic contributions to the binding stability
  of {Barnase and Barstar}.
\newblock Biophys. J. \textbf{85}(1), 49--60 (2003).
\newblock \urlprefix\url{http://www.biophysj.org/cgi/content/abstract/85/1/49}

\bibitem{Duan:2001}
Duan, Z.H., Krasny, R.: An adaptive treecode for computing nonbonded potential
  energy in classical molecular systems.
\newblock J. Comput. Chem. \textbf{22}(2), 184--195 (2001).
\newblock \doi{10.1002/1096-987X(20010130)22:2<184::AID-JCC6>3.0.CO;2-7}.
\newblock
  \urlprefix\url{http://dx.doi.org/10.1002/1096-987X(20010130)22:2<184::AID-JCC6>3.0.CO;2-7}

\bibitem{Elsen:2006}
Elsen, E., Houston, M., Vishal, V., Darve, E., Hanrahan, P., Pande, V.: {N-Body
  Simulation on GPUs}.
\newblock In: Proceedings of the 2006 ACM/IEEE Conference on Supercomputing, SC
  '06, pp. 188--es. Association for Computing Machinery, New York, NY, USA
  (2006).
\newblock \doi{10.1145/1188455.1188649}.
\newblock \urlprefix\url{https://doi.org/10.1145/1188455.1188649}

\bibitem{Geng:2013}
Geng, W.: Parallel higher-order boundary integral electrostatics computation on
  molecular surfaces with curved triangulation.
\newblock J. Comput. Phys. \textbf{241}, 253 -- 265 (2013).
\newblock \doi{http://dx.doi.org/10.1016/j.jcp.2013.01.029}.
\newblock
  \urlprefix\url{http://www.sciencedirect.com/science/article/pii/S0021999113000739}

\bibitem{Geng:2015}
Geng, W.: A boundary integral {Poisson--Boltzmann} solvers package for solvated
  bimolecular simulations.
\newblock Computational and Mathematical Biophysics \textbf{3}, 43--58 (2015)

\bibitem{Geng:2013a}
Geng, W., Jacob, F.: {A GPU-accelerated direct-sum boundary integral
  Poisson-Boltzmann solver}.
\newblock Comput. Phys. Commun. \textbf{184}(6), 1490 -- 1496 (2013).
\newblock \doi{http://dx.doi.org/10.1016/j.cpc.2013.01.017}.
\newblock
  \urlprefix\url{http://www.sciencedirect.com/science/article/pii/S0010465513000349}

\bibitem{Geng:2013b}
Geng, W., Krasny, R.: A treecode-accelerated boundary integral
  {Poisson-Boltzmann} solver for electrostatics of solvated biomolecules.
\newblock J. Comput. Phys. \textbf{247}, 62 -- 78 (2013).
\newblock \doi{http://dx.doi.org/10.1016/j.jcp.2013.03.056}.
\newblock
  \urlprefix\url{http://www.sciencedirect.com/science/article/pii/S0021999113002404}

\bibitem{Geng:2007}
Geng, W., Yu, S., Wei, G.W.: Treatment of charge singularities in implicit
  solvent models.
\newblock J. Chem. Phys. \textbf{127}, 114106 (2007)

\bibitem{Geng:2017}
Geng, W., Zhao, S.: {A two-component Matched Interface and Boundary (MIB)
  regularization for charge singularity in implicit solvation}.
\newblock J. Comput. Phys. \textbf{351}, 25--39 (2017)

\bibitem{Greengard:1987}
Greengard, L., Rokhlin, V.: A fast algorithm for particle simulations.
\newblock J. Comput. Phys. \textbf{73}(2), 325 -- 348 (1987).
\newblock \doi{http://dx.doi.org/10.1016/0021-9991(87)90140-9}.
\newblock
  \urlprefix\url{http://www.sciencedirect.com/science/article/pii/0021999187901409}

\bibitem{Hamada:2009}
Hamada, T., Nitadori, K., Benkrid, K., Ohno, Y., Morimoto, G., Masada, T.,
  Shibata, Y., Oguri, K., Taiji, M.: {A novel multiple-walk parallel algorithm
  for the Barnes--Hut treecode on GPUs -- towards cost effective, high
  performance N-body simulation}.
\newblock Computer Science - Research and Development \textbf{24}(1), 21--31
  (2009).
\newblock \doi{10.1007/s00450-009-0089-1}.
\newblock \urlprefix\url{https://doi.org/10.1007/s00450-009-0089-1}

\bibitem{Holst:1994}
Holst, M.J.: The poisson-boltzmann equation: Analysis and multilevel numerical
  solution.
\newblock Ph.D. thesis, UIUC (1994)

\bibitem{Huang:2012}
Huang, N., Chelliah, Y., Shan, Y., Taylor, C.A., Yoo, S.H., Partch, C., Green,
  C.B., Zhang, H., Takahashi, J.S.: Crystal structure of the heterodimeric
  {CLOCK:BMAL1} transcriptional activator complex.
\newblock Science \textbf{337}(6091), 189--194 (2012).
\newblock \doi{10.1126/science.1222804}.
\newblock
  \urlprefix\url{http://www.sciencemag.org/content/337/6091/189.abstract}

\bibitem{Humphrey:1996}
Humphrey, W., Dalke, A., Schulten, K.: {VMD} -- visual molecular dynamics.
\newblock J. Mol. Graphics \textbf{14}(1), 33--38 (1996).
\newblock \urlprefix\url{http://dx.doi.org/10.1016/0263-7855(96)00018-5}

\bibitem{Juffer:1991}
Juffer, A., E., B., van Keulen, B., van~der Ploeg, A., Berendsen, H.: The
  electric potential of a macromolecule in a solvent: a fundamental approach.
\newblock J. Comput. Phys. \textbf{97}, 144--171 (1991)

\bibitem{Kang:2021}
Kang, S., Yang, M., He, S., Wang, Y., Chen, X., Chen, Y.Q., Hong, Z., Liu, J.,
  Jiang, G., Chen, Q., Zhou, Z., Zhou, Z., Huang, Z., Huang, X., He, H., Zheng,
  W., Liao, H.X., Xiao, F., Shan, H., Chen, S.: {A SARS-CoV-2 antibody curbs
  viral nucleocapsid protein-induced complement hyperactivation}.
\newblock Nature Communications \textbf{12}(1), 2697 (2021)

\bibitem{Li:2009}
Li, P., Johnston, H., Krasny, R.: A {Cartesian} treecode for screened {Coulomb}
  interactions.
\newblock J. Comput. Phys. \textbf{228}, 3858--3868 (2009).
\newblock \doi{10.1016/j.jcp.2009.02.022}.
\newblock \urlprefix\url{http://dx.doi.org/10.1016/j.jcp.2009.02.022}

\bibitem{Lindsay:2001}
Lindsay, K., Krasny, R.: A particle method and adaptive treecode for vortex
  sheet motion in three-dimensional flow.
\newblock J. Comput. Phys. \textbf{172}(2), 879 -- 907 (2001).
\newblock \doi{http://dx.doi.org/10.1006/jcph.2001.6862}.
\newblock
  \urlprefix\url{http://www.sciencedirect.com/science/article/pii/S0021999101968627}

\bibitem{Lu:2007}
Lu, B., Cheng, X., McCammon, J.A.: A
  new-version-fast-multipole-method-accelerated electrostatic calculations in
  biomolecular systems.
\newblock J. Comput. Phys. \textbf{226}(2), 1348 -- 1366 (2007).
\newblock \doi{http://dx.doi.org/10.1016/j.jcp.2007.05.026}.
\newblock
  \urlprefix\url{http://www.sciencedirect.com/science/article/pii/S0021999107002379}

\bibitem{Lu:2007a}
Lu, B., McCammon, J.A.: Improved boundary element methods for
  {Poisson-Boltzmann} electrostatic potential and force calculations.
\newblock J. Chem. Theory Comput. \textbf{3}, 1134--1142 (2007).
\newblock \doi{10.1021/ct700001x}.
\newblock \urlprefix\url{http://dx.doi.org/10.1021/ct700001x}.
\newblock PMID: 26627432

\bibitem{Nguyen:2017}
Nguyen, D.D., Wang, B., Wei, G.W.: Accurate, robust, and reliable calculations
  of {Poisson-Boltzmann} binding energies.
\newblock J. Comput. Chem. \textbf{38}(13), 941--948 (2017).
\newblock \doi{10.1002/jcc.24757}.
\newblock \urlprefix\url{http://dx.doi.org/10.1002/jcc.24757}

\bibitem{Nyland:2009}
Nyland, L., Harris, M., Prins, J.: Fast N-body simulation with CUDA, \emph{GPU
  Gems}, vol.~3.
\newblock Upper Saddle River, NJ: Addison-Wesley (2009)

\bibitem{Saad:1986}
Saad, Y., Schultz, M.: {GMRES:} a generalized minimal residual algorithm for
  solving nonsymmetric linear systems.
\newblock SIAM J. Sci. Stat. Comput. \textbf{7}, 856--869 (1986).
\newblock \doi{10.1137/0907058}.
\newblock \urlprefix\url{http://dx.doi.org/10.1137/0907058}

\bibitem{Salmon:1986}
Salmon, J.K., Warren, M.S., Winckelmans, G.S.: Fast parallel tree codes for
  gravitational and fluid dynamical n-body problems.
\newblock Int. J. Supercomputer Appl \textbf{8}, 129--142 (1986)

\bibitem{Sanner:1996}
Sanner, M.F., Olson, A.J., Spehner, J.C.: {REDUCED SURFACE}: An efficient way
  to compute molecular surfaces.
\newblock Biopolymers \textbf{38}, 305--320 (1996)

\bibitem{Simonson:2002}
Simonson, T., Archontis, G., Karplus, M.: Free energy simulations come of age:
  Protein-ligand recognition.
\newblock Acc. Chem. Res. \textbf{35}(6), 430--437 (2002).
\newblock \doi{10.1021/ar010030m}.
\newblock \urlprefix\url{http://dx.doi.org/10.1021/ar010030m}.
\newblock PMID: 12069628

\bibitem{Unwin:2005}
Unwin, N.: Refined structure of the nicotinic acetylcholine receptor at {4\AA}~
  resolution.
\newblock J. Mol. Biol. \textbf{346}(4), 967 -- 989 (2005).
\newblock \doi{http://dx.doi.org/10.1016/j.jmb.2004.12.031}.
\newblock
  \urlprefix\url{http://www.sciencedirect.com/science/article/pii/S0022283604016018}

\bibitem{Vandervaart:2023}
Vandervaart, J.P., Inniss, N.L., Ling-Hu, T., Minasov, G., Wiersum, G.,
  Rosas-Lemus, M., Shuvalova, L., Achenbach, C.J., Hultquist, J.F., Satchell,
  K.J.F., Bachta, K.E.R.: {Serodominant SARS-CoV-2 Nucleocapsid Peptides Map to
  Unstructured Protein Regions}.
\newblock Microbiology Spectrum \textbf{11}(3), e00324--23 (2023)

\bibitem{Vaughn:2020a}
Vaughn, N., Wilson, L., Krasny, R.: A {GPU}-accelerated barycentric {L}agrange
  treecode.
\newblock In: 2020 IEEE International Parallel and Distributed Processing
  Symposium Workshop (IPDPSW), pp. 701--710 (2020)

\bibitem{Wagoner:2006}
Wagoner, J.A., Baker, N.A.: Assessing implicit models for nonpolar mean
  solvation forces: The importance of dispersion and volume terms.
\newblock Proceedings of the National Academy of Sciences \textbf{103}(22),
  8331--8336 (2006).
\newblock \doi{10.1073/pnas.0600118103}.
\newblock \urlprefix\url{http://www.pnas.org/content/103/22/8331.abstract}

\bibitem{Wilson:2022a}
Wilson, L., Geng, W., Krasny, R.: {TABI-PB 2.0: An Improved Version of the
  Treecode-Accelerated Boundary Integral Poisson-Boltzmann Solver}.
\newblock The Journal of Physical Chemistry B \textbf{126}(37), 7104--7113
  (2022).
\newblock \doi{10.1021/acs.jpcb.2c04604}.
\newblock \urlprefix\url{https://doi.org/10.1021/acs.jpcb.2c04604}

\bibitem{Wilson:2022}
Wilson, L., Hu, J., Chen, J., Krasny, R., Geng, W.: {Computing electrostatic
  binding energy with the TABI Poisson--Boltzmann solver}.
\newblock Communications in Information and Systems \textbf{22}(2), 247--273
  (2022)

\bibitem{Wilson:2021}
Wilson, L., Vaughn, N., Krasny, R.: {A GPU-accelerated fast multipole method
  based on barycentric Lagrange interpolation and dual tree traversal}.
\newblock Computer Physics Communications \textbf{265}, 108017 (2021)

\bibitem{Zhang:2015}
Zhang, B., Peng, B., Huang, J., Pitsianis, N.P., Sun, X., Lu, B.: Parallel
  {AFMPB} solver with automatic surface meshing for calculation of molecular
  solvation free energy.
\newblock Comput. Phys. Commun. \textbf{190}, 173 -- 181 (2015).
\newblock \doi{http://dx.doi.org/10.1016/j.cpc.2014.12.022}.
\newblock
  \urlprefix\url{http://www.sciencedirect.com/science/article/pii/S0010465515000089}

\bibitem{Zhou:2010}
Zhou, Y.C., Lu, B., Gorfe, A.A.: Continuum electromechanical modeling of
  protein-membrane interactions.
\newblock Phys. Rev. E \textbf{82}, 041923 (2010).
\newblock \doi{10.1103/PhysRevE.82.041923}.
\newblock \urlprefix\url{https://link.aps.org/doi/10.1103/PhysRevE.82.041923}

\end{thebibliography}

\end{document}